\documentclass[]{aa}  

\usepackage{graphicx}
\usepackage{txfonts}
\usepackage{linenoaa}

\newcommand{\galex}{{GALEX}}
\newcommand{\gaia}{{{Gaia}}}

\newcommand{\nuv}{{\rm{NUV}}}
\newcommand{\g}{{\rm{G}}}
\newcommand{\bp}{{\rm{BP}}}
\newcommand{\rp}{{\rm{RP}}}
\newcommand{\teff}{{T_{\rm{eff}}}}

\newcommand{\chired}{{ \chi^2_{\rm{red}} }}
\newcommand{\Lwd}{{L_{\rm{WD}}}}
\newcommand{\lbol}{{L_{\rm{bol}}}}

\defcitealias{Rebassa2021}{RM21}
\defcitealias{nayak_2024_wdms}{N24}
\defcitealias{Rebassa_2025_mag_limited}{RM25}

\usepackage{hyperref} 
\begin{document} 

   \title{Revealing Unresolved White Dwarf–Main Sequence Binaries using Gaia DR3 and GALEX}

   \subtitle{I. A Volume limited study of 100 pc}

   \author{Prasanta K. Nayak
          \inst{1}
          }

   \institute{Instituto de Astrofísica, Pontificia Universidad Católica de Chile, Av. Vicuña MacKenna 4860, 7820436, Santiago, Chile\\
              \email{nayakphy@gmail.com, pnayak@astro.puc.cl}
             }

   \date{Accepted on 20/03/2026}

 
  \abstract
   {Understanding the demographics of white dwarf--main sequence (WDMS) binaries is key to uncovering the formation of various stellar exotica and refining the details of binary stellar evolution. Despite several dedicated efforts to identify unresolved WD–MS binaries, their population remains incomplete, even within a 100 pc volume-limited sample.}
   {This study aims to identify WD–MS binaries hidden within the main sequence of the optical color-magnitude diagram (CMD), improving the completeness of WD–MS binaries within a volume-limited sample of 100 pc. }
   {We use NUV-optical CMDs to distinguish unresolved WD–MS binaries from the rest of the populations. High-precision astrometric and photometric data from \gaia\ DR3 and NUV data from \galex\ GR6/7 are combined to construct CMDs. Using the binary spectral energy distribution (SED) fitting algorithm within the Virtual Observatory SED Analyzer (VOSA) tool, we estimate stellar parameters such as effective temperature, bolometric luminosity, and radii. The WD masses are determined using white dwarf evolutionary models. As we use the sources which are detected only in NUV band of GALEX, this study directly complements to majority of the previous studies.
   }
   {We identify 347 WDMS binary candidates within 100 pc, with 188 newly reported. Our method predominantly identifies binaries having cooler WDs ( $\le$10,000 K) compared to previous studies. The WD masses range from $\sim$0.2 and 1.3 M$_\odot$, and most MS companions are of M spectral type. 
   }
   { }


   \maketitle
   \nolinenumbers

%

\section{Introduction} \label{sec:intro}

A significant fraction of stellar sources form as components of binary or multiple systems \citep{Iben_1991_binary_pop}. Depending on the initial separation in a binary system, the massive star can significantly influence its companion's evolution.
Thus, studying binary stars plays a key role in understanding stellar evolution. However, the majority of them are photometrically unresolved binaries. 
In this context, multi-wavelength approaches helped to identify several white dwarf -- main sequence (WDMS) binaries \citep{Rebassa2012a, Ren2018_wdms_LAMOST_DR5, Anguiano_2022}, yet a long way to reach the completeness even within the solar neighborhoods \citep[][hereafter \citetalias{Rebassa2021}]{Rebassa2021}. A large and homogeneous sample of WD–MS binaries is also essential for informing binary population synthesis studies, providing meaningful constraints on several uncertain aspects of binary stellar evolution \citep[][]{Toonen2017,torres_2022}.

Investigating compact WDMS binaries, where the systems go through a common envelope (CE) phase, is key to constraining the efficiency and energy budget of CE evolution \citep[][]{Zorotovic2010, Rebassa2012b, Camacho2014, Zorotovic2014}, determining the critical mass ratio that governs stable versus unstable mass transfer \citep[e.g.][]{Bobrick_2017}, and refining the mass-radius relation of white dwarfs \citep[][]{Parsons2017}. 
In close compact binaries, mass accretion onto a WD can also potentially trigger a Type Ia supernova, a key astronomical event \citep[][]{Wang2012}. 
Conversely, mass loss from the WD progenitor, through either CE or Roche lobe overflow (RLOF), can lead to the formation of extremely low-mass white dwarfs (ELM-WDs) with masses $\le$ 0.3 M$_\odot$ \citep{istrate_2014_a,istrate_2014_b, nandez_2015}. 
On the other hand, in wide-orbit WD–MS binaries, both stars evolve independently without influencing each other's evolution, leading the more massive star to eventually become a white dwarf \citep[][]{Garcia1997, Farihi2010_WD-RD}. Analyzing these wide systems provides valuable insights into the age-metallicity relation in the solar neighborhood \citep[][]{Rebassa2016b, Rebassa2021}, the secondary mass function \citep[][]{Ferrario2012}, and the relationship between age, activity, and rotation \citep[][]{morgan2012, Rebassa2013a, Skinner2017}.
Therefore, studying WDMS binaries is essential for uncovering the formation of diverse stellar exotica and constraining the processes involved in binary stellar evolution.

There have been several previous attempts to search for a large number of WDMS binaries using various tools like UV color-temperature relation,  UV-optical color-magnitude diagrams and using $\chi^2$ template-fitting techniques or wavelet-based spectral decomposition on observed spectra, where the hot WD contributes distinct H-Balmer lines or continuum features in the blue/UV region. A few thousand spectroscopic WDMS binaries with M-dwarf companions are detected using Sloan Digital Sky Survey \citep[SDSS;][]{York2000_sdss, Eisenstein2011_sdssIII, Rebassa2012a, Rebassa2016a} and the Large Sky Area Multi-Object Fiber Spectroscopic Telescope Survey \citep[LAMOST;][]{Cui2012_LAMOST, Chen2012_LEGUE_LAMOST,  Ren2018_wdms_LAMOST_DR5}. The \gaia\ DR3's low resolution BP/RP (XP) spectra have also been recently used to identify WDMS binaries \citep{Rebassa_2025_mag_limited, Xabier_2025_ML_WDMS, Li_2025_30k_wdms}. Several previous studies \citep[][]{WDS1_parsons2016, WDS2_Rebassa2017, Anguiano2020_wd_APOGEE, WDS5_Ren2020} used $\teff$ vs (FUV$-$NUV) color relation to distinguish WDMS binaries from the MS populations, where $\teff$ values are obtained from LAMOST or Radial Velocity Experiment \citep[RAVE;][]{Kordopatis2013_RAVE_DR4, Kunder2017_RAVE_DR5} and UV data is obtained from Galaxy Evolution Explorer \citep[GALEX;][]{martin2005_GALEX}. Recently, \citet{nayak_2024_wdms} and \citet{Jackim_2024_WDMS_galex} used UV and optical CMDs as a tool to separate WDMS binary candidates from MS populations by combining \gaia\ and \galex\ data. Astrometric solutions from the Tycho-\gaia\ astrometric solution \citep[TGAS;][]{Michalik2015_TGAS}, \gaia-DR2 \citep[][]{Evans2018_gaia_DR2_photometry}, and \gaia\ DR3 \citep[][]{Gaia_DR3_NSS} have also been used to identify WDMS candidates \citep{Andrews2022, Shahaf2023}.

Despite several dedicated studies in identifying unresolved WDMS binaries, their population is still incomplete even within a volume-limited sample of 100 pc \citepalias[][]{Rebassa2021}. Through simulation, \citetalias{Rebassa2021} showed that $\approx91\%$ of WDMS binaries are hidden within the MS region of optical CMD, however, their study focused on the remaining $\approx9\%$ of the WDMS binary populations within 100 pc volume of the solar neighborhood located in the gap region between WD and MS regions in optical CMD. 
As mentioned above, spectroscopic surveys in combination with UV photometric surveys were also used to find these hidden WDMS binaries in the MS but all the current spectroscopic surveys are restricted by magnitude-limit. The binary candidates found by astrometric solutions also require further UV or spectroscopic observations to confirm their candidature \citep{Ganguly_2023}. 

Recently, \citet[hereafter \citetalias{nayak_2024_wdms}]{nayak_2024_wdms} targeted to reveal the hidden $\approx91\%$ of WDMS binaries within 100 pc volume. The authors used UV and optical CMDs by combining the data from \galex\ and \gaia\ DR3 to identify the binaries. However, \citetalias{nayak_2024_wdms} consider only those sources having both FUV and NUV detections in \galex\ which leads them to find WDMS binaries with relatively hotter WDs. 
Studies by \citetalias{Rebassa2021} and 
\citet[hereafter \citetalias{Rebassa_2025_mag_limited}]{Rebassa_2025_mag_limited} included sources where only NUV observation or no \galex\ UV observations are present, however, the sources are confined within the gap region between WD and MS in the \gaia\ CMD. 
In a recent study on WDs and WDs in binaries using \gaia\ and \galex\ , \cite{Jackim_2024_WDMS_galex} also included the sources which are only detected in the NUV band. However, there are a few things which were not considered in their study and can be included for better estimation of WDMS parameters. The artifacts and spurious sources from the \galex\ data were not removed. Primary selection of WD binary candidates using UV --optical CMDs does not include extinction correction. Though extinction was included in determining the parameters of WD binary candidates, the use of 2D extinction \citep{Schlegel1998} leads to larger values even for the sources located within a few 100 pc where the extinction is found to be negligible \citepalias[][]{Rebassa2021, nayak_2024_wdms}. The parameter estimation of WD binaries mainly includes the data from \galex\, and \gaia\ , and wherever SDSS data are available. Considering the majority of the MS companions in the binary samples are low-mass M-dwarfs, spectral energy distribution with a larger wavelength coverage including the near-infrared (NIR) observations is necessary for the better estimation of their parameters. Choice of wrong extinction values and non-inclusion of NIR observations can make a WD binary appear as a single WD \citepalias[][]{nayak_2024_wdms}.  
Therefore, better constraints on the selection of data with a longer wavelength coverage and the use of 3D extinction models are necessary to construct a homogeneous and complete sample of WDMS binaries and to improve the estimation of their parameters.

In this paper, we aim to identify the WDMS binaries which are hidden within the MS of the optical CMD within a volume-limited sample of 100 pc to improve the completeness of WDMS binaries. Within 100 pc, photometric and astrometric solutions are expected to be more accurate, and less affected by Galactic extinction. We use \gaia\ DR3 and \galex\ GR6/7 data for the initial selection of WDMS candidate binaries and implemented several cut-off criteria to remove artifacts and sources with bad photometry and astrometric solutions. Nevertheless, we decided to primarily focus on the \galex\ sources having no FUV detection or having FUV detection with artifacts for this study, which makes this work complementary to the previous studies \citep{WDS5_Ren2020, Anguiano_2022, nayak_2024_wdms}, where sources with both FUV and NUV detections are considered.
Identification and parameterization of WDMS binaries, where only NUV or no UV detections are available, are also present in the literature but the authors had studied either only the gap region of the \gaia\ CMD \citepalias[][]{Rebassa2021, Rebassa_2025_mag_limited} or only the WD region \citep{Xabier_2025_ML_WDMS}, or used the optical color-color and CMDs \citep{Li_2025_30k_wdms} to identify the WDMS binaries. Therefore, this work also complements the above mentioned studies.  Our selection criteria also include the NUV bright MS sources located in the gap region between WD and MS in the NUV--optical CMD, which were not included in the recent study by \cite{Jackim_2024_WDMS_galex}.  In our sample selection, we also included the sources which have both FUV and NUV detection with good photometric detections as mentioned in \citepalias[][]{nayak_2024_wdms}, but used only NUV data for identification and parameter estimations. Parameters of these sources are compared with the literature values \citepalias[][]{nayak_2024_wdms, Rebassa2021, Rebassa_2025_mag_limited} and \citep{WDS5_Ren2020}, 
where both FUV and NUV data were used for parameter estimation, which gives confidence in our estimates and in implementing our method to the sources with only NUV detection. 
Finally, we examine the possibility of contamination from active M-dwarfs in our catalog, which are also known for their excess UV emission due to stellar activity. However, the literature suggests that such contamination is found to be negligible in the UV and UV--optical CMDs \citep{Anguiano_2022, nayak_2024_wdms, Jackim_2024_WDMS_galex}.

In \autoref{sec:methods}, we describe the data selection criteria, the methods used to identify the WDMS binary candidates and to estimate their stellar parameters. In \autoref{discussion}, we discuss our results and a separate method to estimate WD mass (\autoref{S:wd_mass}, examine the contamination from chromospherically active MS stars in our catalog (\autoref{sec:active_M-dwarf}), compare our catalog with previous studies (\autoref{comparison}), compare with optical spectroscopic surveys (\autoref{sec:spectroscopy} and determine the completeness of our catalog (\autoref{completeness}). We conclude our results in \autoref{conclude}.


\section{Data Selection and Characterization}
\label{sec:methods}
In this section, we describe the cut-off criteria to select reliable \gaia\ DR3 and \galex\ NUV data and the selection procedures of WDMS candidate binaries in this study. Furthermore, we also discuss the estimation of the stellar properties of both companions of these binaries.


\begin{figure*}
	
	\includegraphics[width=2\columnwidth]{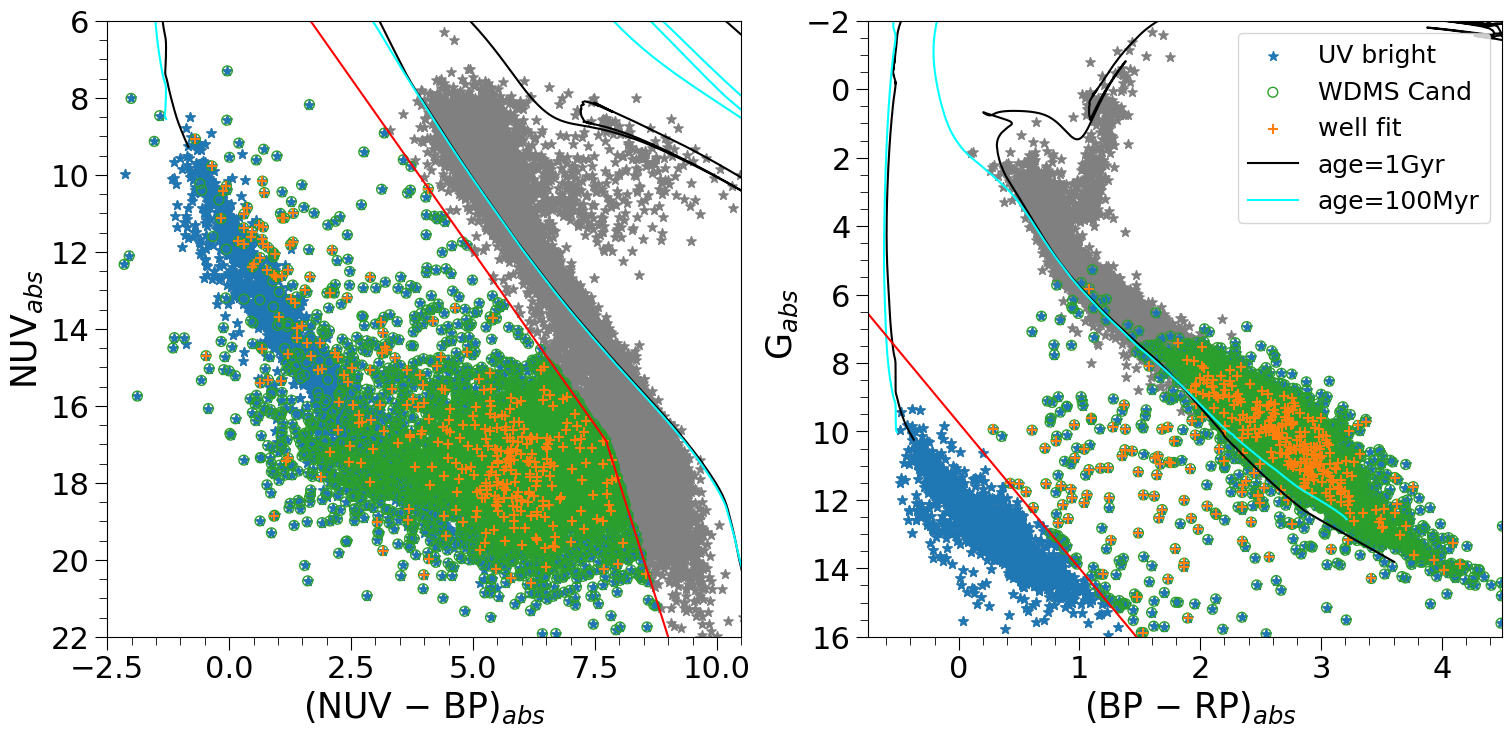}
    \caption{ The CMDs for sources within 100\ pc in the absolute plane after correcting for distance modulus and extinction, where the subscript "abs" stands for absolute magnitude. In both the panel, isochrones of 100 Myr (cyan line) and 1 Gyr (black line) are overlaid to indicate the evolutionary sequences, and the grey asterisks indicate main-sequence (MS) and post-MS populations.  \\ 
    \textit{Left:} UV-Optical CMD (NUV$-$BP vs NUV). The red line separates the hotter and NUV-bright sources (blue asterisks, green circles and orange plus) from the others (grey asterisks) on the NUV-optical CMD. \\
    \textit{Right}: Optical CMD (BP$-$RP vs G). The red line separates the blue asterisks (WD populations) from the rest of the NUV-bright sources (green circles and orange plus) which are located on the MS or with within the gap region between WD and MS. These green circles represent WDMS binary candidates. Orange plus highlights the candidates for which we could fit observed SED from UV to IR with WDMS composite model fluxes using the VOSA tool.
     }
    \label{selection}
\end{figure*}

\begin{table}[]
    \centering
    \caption{Summarize the number of sources after each data selection criteria.}
    \begin{tabular}{|c|c|c|c|}
    \hline
       Selection Criteria  & \multicolumn{3}{c|}{{no. of sources}} \\
       \cline{2-4}
        & combined & FUV+NUV & only NUV \\
       \hline
        100 pc Gaia DR3  & 574531 & --- & --- \\
        astrometric cuts & 306515  & --- & --- \\
        GALEX match & 50985   & 15098 & 35887 \\
        cuts on GALEX  & 30486   & 7355 & 23131 \\
        NUV bright & 9751   & 2448 & 7303 \\
        WDMS candidates & 6559   & 1254 & 5305 \\
         well-fitted SEDs & 347  & 100 & 247 \\
        New WDMS  & 188  & 15 & 173 \\
        \hline
    \end{tabular}
    \label{tab:data_selection}
\end{table}


\subsection{Data Selection} 
\label{data}

First, we use \gaia\ DR3 to identify the sources within 100 pc. Then, we applied the following conditions to remove the sources with larger photometric (fluxes in G, BP and RP bands) and astrometric (parallax) errors \footnote{Cut-offs in the renormalized unit weight error (RUWE) and astrometric excess noise (AEN) are often used in the literature for source selection. We intentionally avoid using these parameters since it has been shown that binarity may lead to high values of RUWE and AEN \citep[][]{Belokurov2020}. } : 

\begin{itemize}
\item $\varpi/\sigma_{\varpi} \ge$ 10 
\item $I_{\rm BP}/\sigma_{I_{\rm BP}} \ge$ 10
\item $I_{\rm RP}/\sigma_{I_{\rm RP}} \ge$ 10
\item $I_{\rm G}/\sigma_{I_{\rm G}} \ge$ 10
\end{itemize}

\noindent where, $\varpi$  is the parallax in  arcseconds, $I_{\rm G}$, $I_{\rm BP}$ and  $I_{\rm RP}$ are the fluxes in  the bandpass filters $G$, ${BP}$ and ${RP}$, respectively, and $\sigma$ are the standard errors of the corresponding parameters. We find 306,515 \gaia\ sources after applying the above selection criteria. \gaia\ provides source coordinates in J2016.5 epoch, whereas all the other archival UV, optical and NIR datasets are cataloged in J2000. Therefore, we used \gaia\ DR3's astrometric solutions to convert the coordinates from J2016.5 to J2000 epoch. 
We then cross-matched these sources with the \galex\ GR6+7 catalog and found 50,985 sources in common (after removing duplicates matches) detected either in both the FUV and NUV bands or at least in the NUV. 
To ensure reliable photometry, we  applied quality cuts to the \galex\ data, retaining only point-like sources detected in the NUV band. Specifically, we selected sources flagged as Nafl = 0 and Nexf = 0, thereby removing artifacts (e.g., bright star near field edge, bad pixel etc.) and extended sources appeared in the NUV band of the \galex\ catalog. \footnote{http://vizier.cds.unistra.fr/viz-bin/VizieR-3?-source=II/312}
We did not impose any quality cuts on the FUV detections, as the FUV data were not used for identification or parameterization in the present analysis. Thus, the sample includes sources with both good and poor FUV data but reliable NUV measurements. Identification of WDMS binary candidates having reliable photometry in both FUV and NUV bands has already been studied by \citetalias{nayak_2024_wdms}, therefore, this study directly complements that work. 
We notice that all the selected sources have NUV photometric errors $\le$0.5 mags, so, we did not apply additional filtering criteria on NUV photometric errors. 
Applying these source selection criteria, we are left with 30,486 sources. The summary of data selection is presented in \autoref{tab:data_selection}.


\subsection{Identification of Candidates} 
\label{sec:identification}
We analyzed these 30,486 sources to construct optical and NUV$-$optical CMDs in absolute magnitudes plane after taking into account the distance modulus obtained from \gaia's DR3 catalog \citep[][]{bailer-jones_2021edr3} and extinctions obtained from a three-dimensional (3D) Dust Map 
\citep[MWDUST \footnote{https://github.com/jobovy/mwdust};][]{Drimmel_2003_mwdust, Marshall_2006_mwdust, Bovy_2016_mwdust, Green_2019_mwdust}.
We used these CMDs as a tool to identify WDMS binary candidates. 
The left panel of \autoref{selection} shows the NUV-optical CMD (NUV$-$BP vs NUV) of all the sources of interest. Cyan and black lines represent isochrones of ages 100 Myr and 1 Gyr, respectively, generated using the \textbf{MIST} isochrone package \citep{mist0, mist1} to demonstrate the MS and post-MS (grey asterisks), and WD evolutionary sequences as references. The blue asterisks and green circles both indicate the NUV bright populations which include WD and the sources fall in the gap region between the WD and MS populations, separated by the red lines using the following photometric cuts:
\begin{itemize}
\item  $\nuv_{abs} > 1.8\times(\nuv-{\rm BP})_{abs} + 3.0$
\item  $\nuv_{abs} > 4.0\times(\nuv-{\rm BP})_{abs} - 14.0$
\end{itemize}
Now we mark these NUV bright sources in the optical CMD in the right panel of \autoref{selection} to check whether they are distributed in similar evolutionary sequences or not. 
We noticed that the hot NUV-bright population identified in the NUV$-$optical CMD divided into two distinct groups in the optical CMD separated by the red line, which is defined as $\g_{abs} = 4.25\times(\bp -\rp)_{abs} + 9.75$. The blue asterisks represent the WD populations, while a large fraction ($\sim$65\%) of the NUV-bright sources marked as green circles show up in the MS region and a few in the gap region between the MS and the WD cooling sequence. 
This suggests that the sources marked as green are the potential unresolved WDMS binary candidates. When the WD mainly contributes to the UV and the MS to the optical, the source appears among the hot WDs and in the gap region in the NUV$-$optical CMD, but shows up among the MS stars in the optical CMD. However, if the WD also contributes to the optical, the source might appear in the gap region between the MS and the WD cooling sequence in the optical CMD. We identified 6,559 candidates based on this shift in CMDs within our sources of interest.


\begin{figure*}
	\includegraphics[width=0.950\columnwidth]{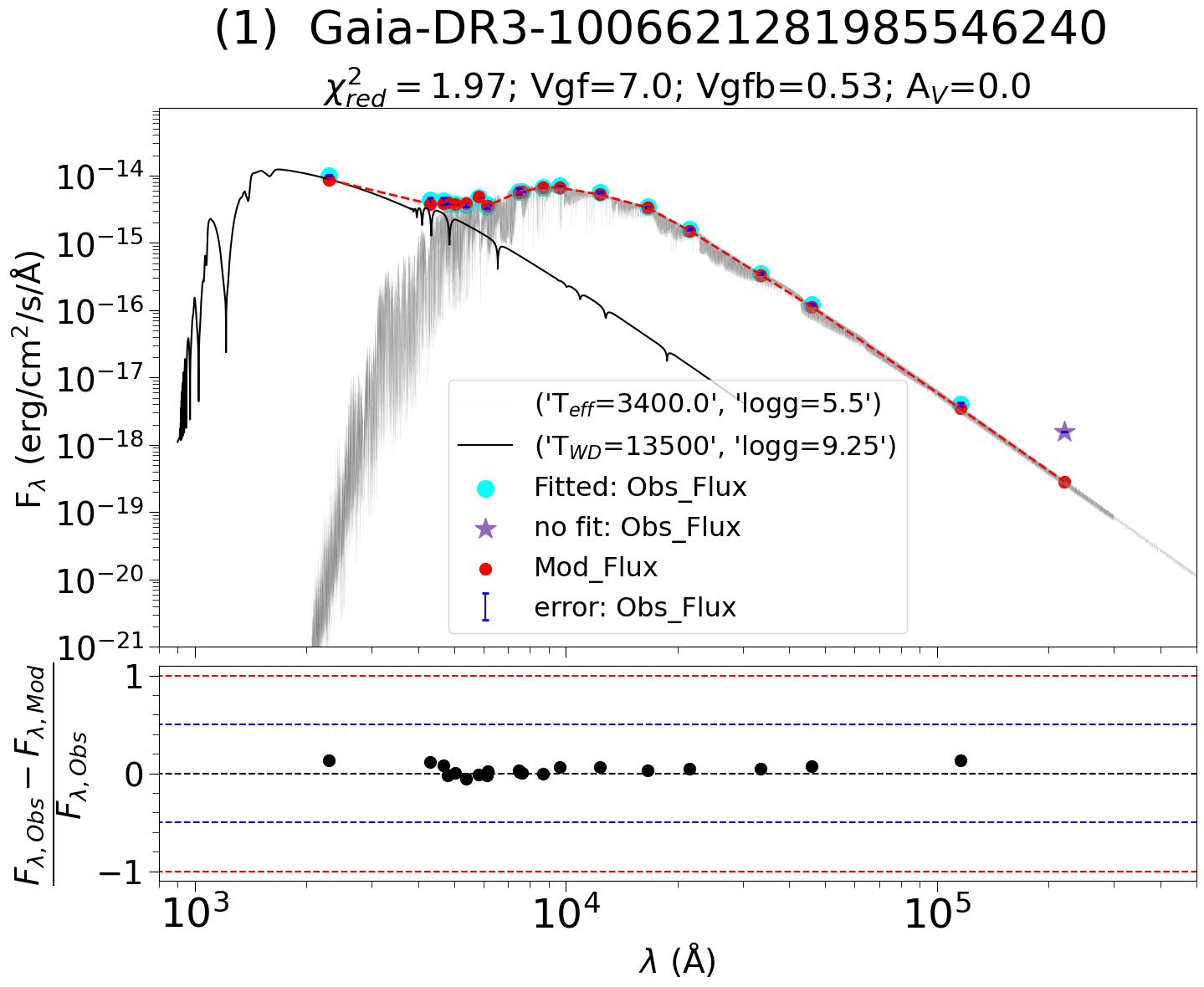} \includegraphics[width=0.950\columnwidth]{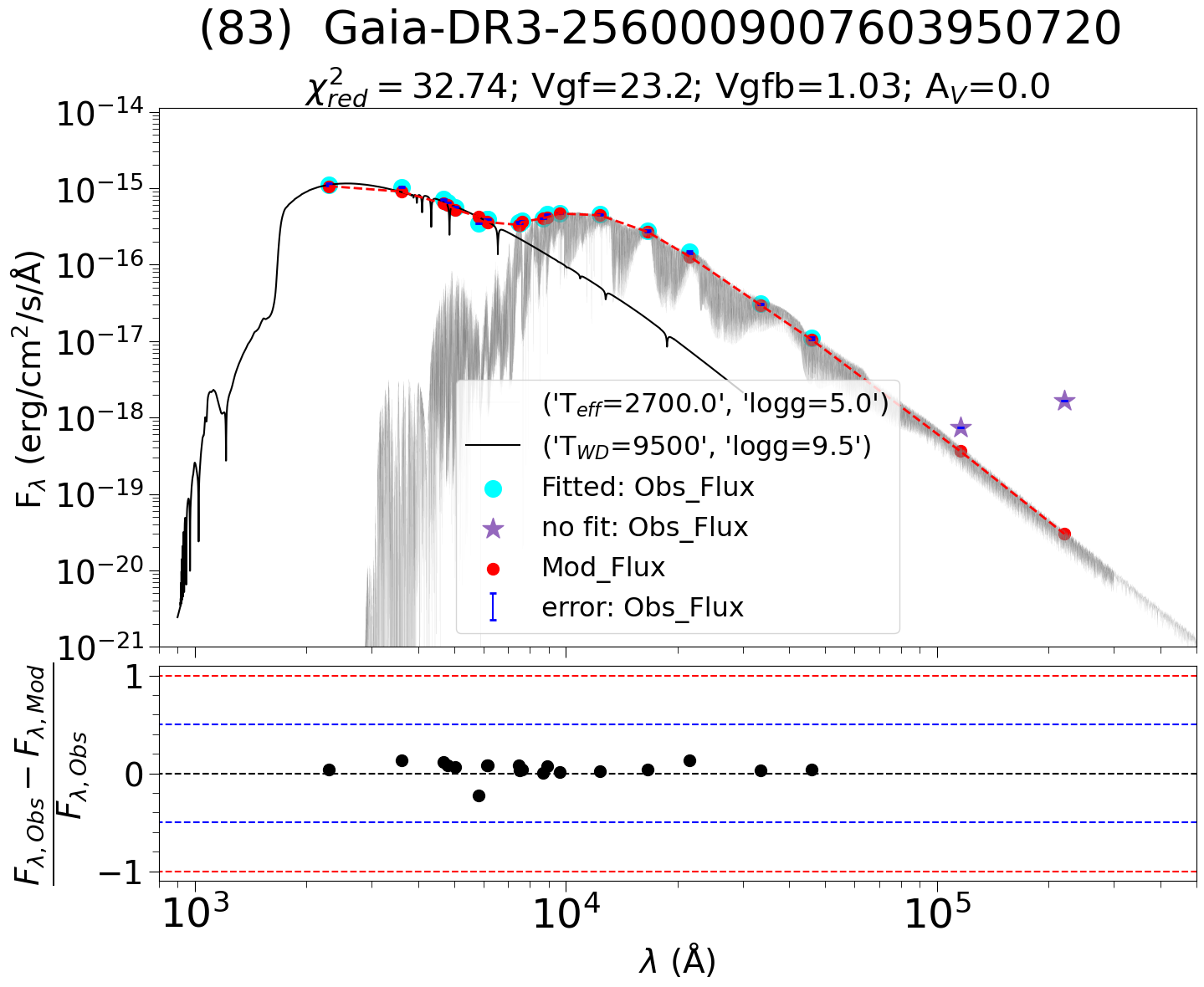} 
    \caption{Two examples of well-fitted SEDs of WDMS binaries. Index number as per \autoref{tab:source_table}, \gaia\ DR3 source ID, the values of A$_V$, $\chired$, Vgf and Vgf$_b$ are mentioned on top of each panel. (Top panel:) Cyan points (with blue errors) denote the observed flux from UV to IR. The observed data points with upper-limit on flux or unreliable detections are marked as asterisks, and they are not included in the fit. The black (grey) line represents the best-fit synthetic spectra of WD (MS). 
    The red points indicate the expected combined model fluxes from the best-fit synthetic spectra. (Bottom panel:) Fractional residue fluxes are shown in different bands. The blue and red dashed lines represent 50\% and 100\% residue flux, while the black line represents zero residue. 
    There is excellent agreement between the cyan and red points, and less than 50\% residue fluxes in every band indicating a well-fitted SED. 
    }
    \label{sed}
\end{figure*}

\begin{figure*}
	\includegraphics[width=0.950\columnwidth]{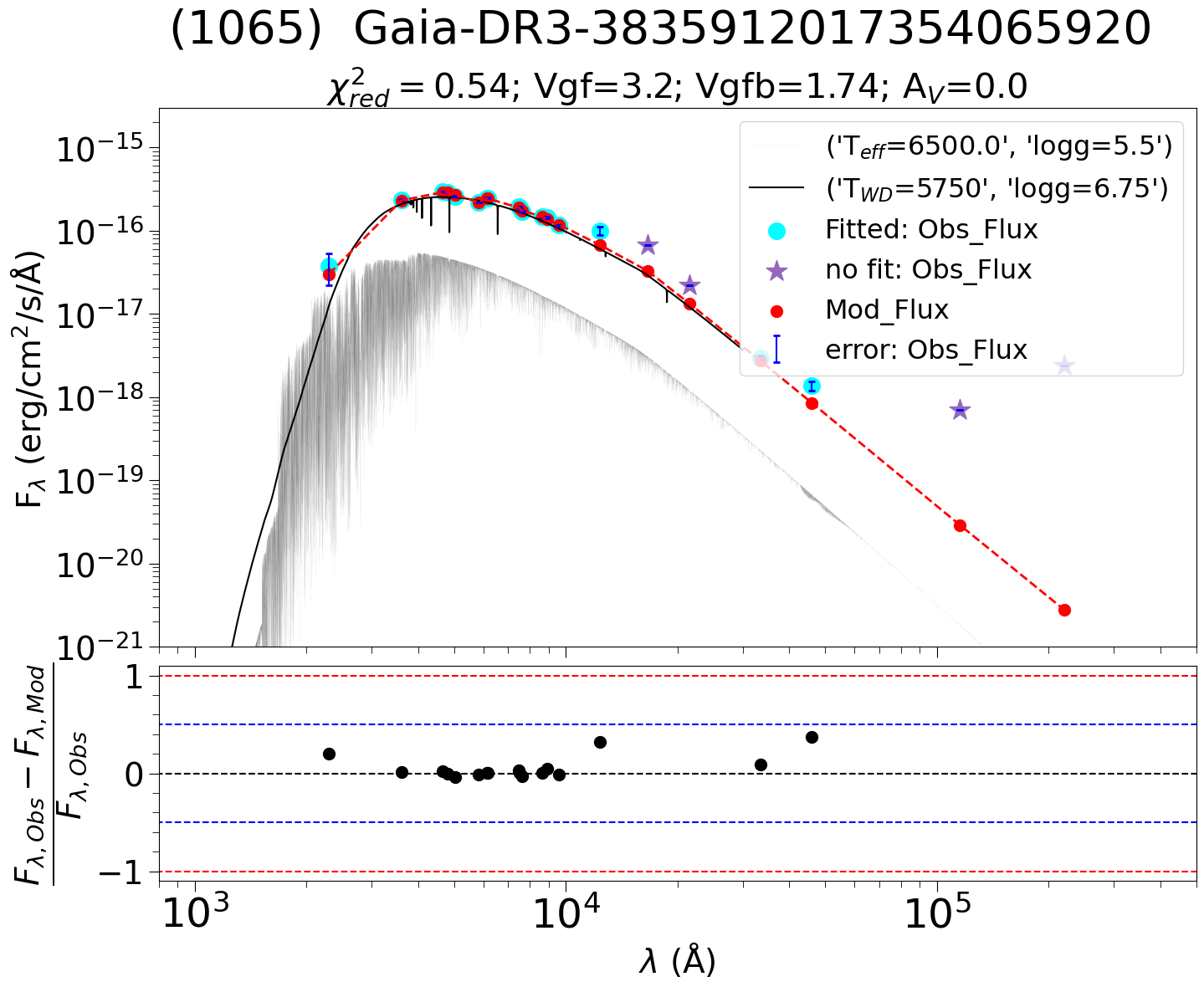} \includegraphics[width=0.950\columnwidth]{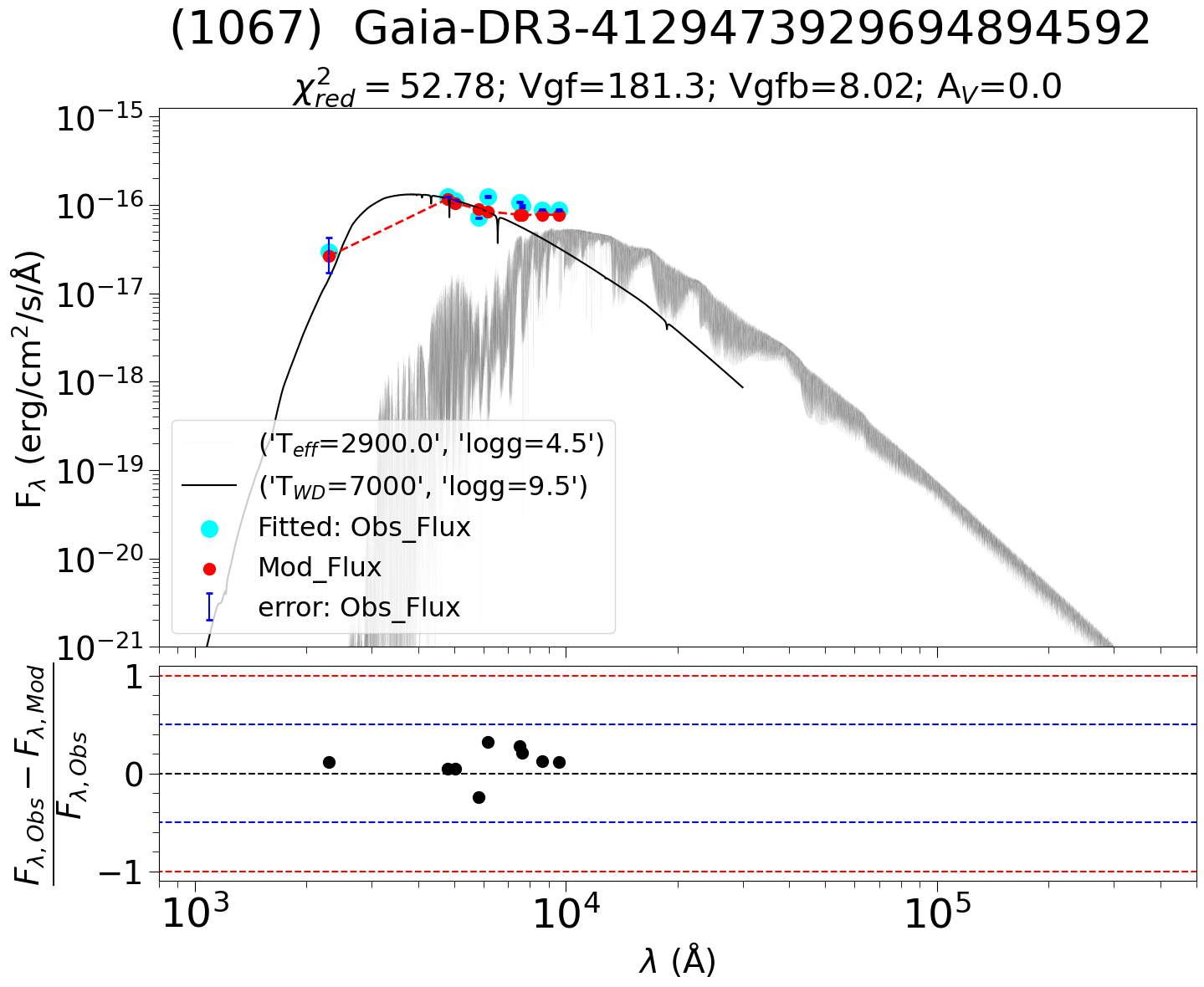} 
    \caption{Same as \autoref{sed} but for two representative examples of badly-fitted SEDs of WDMS candidate binaries. Index number as per the online catalog at the CDS, \gaia\ DR3 source ID, the values of A$_V$, $\chired$, and Vgf$_b$ are mentioned at the top. In the case of source 415 (left), the model spectra are unable to fit the observed flux for majority of data points despite their residual flux $\le$50\% and Vgf$_b$ value $<$ 15. For source 416 (right), though the model spectra nicely fit the observed fluxes in the UV and optical region but only a couple of observed data points are fitted with MS spectrum due to lack of data points in the NIR regions, hence the fit is not reliable. 
}
    \label{badfit}
\end{figure*}

\subsection{Characterization of WDMS Binaries} 
\label{sec:wd_params}

To confirm whether these 6,559 candidates are indeed WDMS binaries and to determine their stellar parameters, we further analyze their spectral energy distributions (SEDs). To construct observed SEDs of the WDMS candidates ranging from UV to NIR wavelengths, we first cross-match them with following surveys: APASS DR9 \citep{apass}, PanSTARRS-DR2 \citep{magnier_panstarrs_dr2} and SDSS DR16 for optical, and 2MASS \citep{2mass} and ALLWISE \citep{allwise} for NIR observations. We did not get any crossmatch for only one source and we proceed with 6558 candidates for SED analyses. We use the virtual observatory SED analyzer's \citep[VOSA;][]{bayo2008} 
in order to construct observational SEDs, to fit the observed SEDs with combined theoretical models of WD and MS using binary fit algorithm and to extract the best-fit stellar parameters for both companions simultaneously.

We use the BT-Settl-CIFIST models \citep{Baraffe2015} for the MS star.   
For our purposes, we use the full range for $\teff$ (1,200 to 7,000 K) available in VOSA but restrict ourselves to $4.5\leq \log\ g\leq5.5$ since MS stars are expected to have $\log\ g$ within this range. 
We use WD evolution models of \citet{Koester2010} included in the VOSA toolkit which provides spectra for H-rich WDs.
We consider the full range of $\teff$ (5,000 to 80,000 K) and $\log\ g$ (6.5 to 9.5) for WD models.
Since our sources are all within 100\,pc by selection, we consider solar metallicity for all of them. 
VOSA performs multiple iterations by varying $\teff$, $\log\ g$, and scaling factor ($M_d$) to minimize $\chi^2$ to find the best-fit spectra to the observed flux distribution. 
The scaling factor $M_d\equiv (R/D)^2$ is used to scale the model flux to match the observed flux, where $R$ and $D$ denote the radius and distance of the source. In our exercise, since the distance is already known from \gaia, the scaling factor gives us the radius.

VOSA provided best-fit stellar parameters ($\teff$, $\log\ g$, luminosity, radius of both the components) using a $\chi^2$ minimization technique  
for 6,556 candidates and it was unable to provide a solution for only two candidates because of the small number of data points. However, unlike in the study of \citetalias{nayak_2024_wdms}, where both FUV and NUV data (representing the WD component of a binary) were used for SED analyses, in the present work only a single data point, the NUV flux, is present in the UV region. Although, the presence of excess emission in the NUV band suggesting that these sources are most probable WDMS binary candidates, fitting WD theoretical model spectra mainly to one UV data point can lead to degenerate or poorly constrained of the WD parameters estimation, particularly when combined with the dominant optical–NIR flux from the main-sequence star. To reduce such degeneracies, 
we introduced a selection step. For each source, we examined the fractional residual flux ($f_{\rm{residue}}$), defined as the fractional difference between the observed and model-predicted flux $((F_{\lambda,\rm{obs}}-F_{\lambda,\rm{model}})/(F_{\lambda,\rm{obs}}))$, where $F_\lambda$ is the observed flux at waveband $\lambda$ and subscripts `obs' (`model') denotes the observed (best-fit model) flux. We selected those candidates for which the $f_{\rm{residue}}$ exceeded 50\% in both the NUV band and in the next bluest optical band with respect to the BT-Settl-CIFIST model spectra representing single main-sequence stars.\footnote{VOSA output provides best-fitted model fluxes for both WD and MS model spectra. It also provides best-fitted combined flux from WD and MS model spectra. In this case, we used F$_{\lambda,\rm{model}}$ as MS model (BT-Settl-CIFIST) spectra for respective bands, NUV and the next bluer optical band to NUV.   
Using this criterion, we identified 1,417 sources that exhibit significant flux excess relative to the single-star MS model. 
These sources likely contain an additional hot component, consistent with a WD and therefore cannot be explained by a single main-sequence star spectrum alone. In this way, we got at least two data points included the WD model spectra fitting. }

However, we note that the implementation of the selection criterion on $f_{\rm{residue}}$ introduces a selection bias toward WDMS binaries hosting cooler white dwarfs and late-type MS companions (i.e., K and M dwarfs). The GALEX NUV band, with an effective wavelength of $\lambda_{\rm eff} \approx 2316~{\AA}$, is primarily sensitive to blackbody temperatures of $\sim$10,000-16,000 K, as inferred from Wien’s displacement law ($\lambda_{\rm max}T = 2.897 \times 10^{-3},\mathrm{m,K}$). 
Consequently, our method is expected to  detect WDs preferentially cooler than $\sim$16,000 K, while the parameter estimation for hotter WDs is expected to carry larger uncertainties due to lack of FUV data point.
In addition, this selection criterion requires that the bluest optical band remains dominated by WD flux, which in turn biases the MS companions toward later spectral types. Considering the bluest optical point to be the Pan-STARRS $g$ band, with $\lambda_{\rm eff} \approx 4810~{\AA}$ (corresponding to blackbody temperatures of $\sim$5,300-7,200 K via Wien’s law), the companion stars in our sample are expected to be cooler than $\sim$5,000 K, corresponding predominantly to K-M spectral types.

As reported in the literature \citepalias[][]{Rebassa2021, nayak_2024_wdms}, smaller observational errors often lead to higher $\chi^2$ values despite the visual inspection suggesting them to be well-fitted SEDs. Conversely, visually inspected badly fitted SEDs can also have smaller $\chi^2$ due to larger photometric errors. Therefore, following \citetalias{Rebassa2021} and \citetalias{ nayak_2024_wdms}, we implemented two criteria for the goodness of the fit:

   \[
      \begin{array}{lp{0.7\linewidth}}
         Vgf_b < 15  & Vgf$_b$ is called the visual goodness of fit, introduced in the VOSA by modifying the $\chired$ formula, where the error is considered to be at least 10\% of its observed flux\footnote{http://svo2.cab.inta-csic.es/theory/vosa/helpw4.php?otype=star\&action=help}.     \\
        \lvert f_{residue} \lvert < 0.5               &  absolute fractional residual flux ($\lvert f_{\rm{residue}} \lvert$) is defined as   $\lvert (F_{\lambda,\rm{obs}}-F_{\lambda,\rm{model}})/(F_{\lambda,\rm{obs}}) \rvert$.  
         Here $F_{\lambda,\rm{model}}$ corresponds to best-fitted combined flux of WD and MS theoretical model spectra at the corresponding waveband $\lambda$. The NUV data and the next bluest optical optical band data must individually satisfy this criterion. In addition, in the optical-IR region, more than 80\% of the data points must satisfy this criterion.                     \\
      \end{array}
   \]
Therefore, we consider the SEDs to be well-fitted by the theoretical models when the sources satisfy both the Vgf$_b$ and the $ f_{residue} $ criteria simultaneously. We find that there are 361 candidates with well-fitted SEDs out of 1,417. The SEDs of the other 1,056 candidates could not be fitted well with WDMS composite model spectra using VOSA according to our criteria. 
The poor fit to the rest of the candidates to the WDMS binary model could be due to various factors, including, but not limited to, inaccurate cross-matching, UV excess originating from stellar activity of the MS star rather than a WD, or contamination from nearby stars in the optical/IR.

In \autoref{sed}, we show the SEDs of two well-fitted WDMS binaries.  Observed fluxes in different filters are represented as cyan points with blue error bars. The asterisks symbols denote bad photometric data points with artifacts or upper-limit flux values, and these data points are not included in the fit.  
The black and grey spectra represent the best-fit WD and MS models to the observed SEDs, respectively. The combined model fluxes in the respective filters are represented by red points and linked with a red dashed line. The best-fitted parameters for WD and MS are noted in the legend. \gaia\ DR3 source ID, $\chired$, Vgf, Vgf$_b$, and A$_V$ values are also noted at the top of each plot. 
$f_{residue}$ values in respective bands are shown in the bottom panel of each plot. The match between the cyan and red points, and $ f_{residue} $ of less than 50\% in all the bands indicate goodness of fits based to our criteria. 
However, after visual inspection of all the well-fitted 361 SEDs, we notice that in some SEDs either UV part is fitted with MS SED model while the optical-NIR regions is fitted with WD SED model which leads to  hotter temperature for MS companion compared to WD (as shown in the left panel of \autoref{badfit}), or temperature difference between the MS and WD component is a few 100 K which is comparable to error in temperature, or only a couple of points used to fit the MS part of the SEDs (as shown in the right panel of \autoref{badfit}), or over-fitted the photometric bands in NIR region.   
We further exclude 14 such sources from well-fitted SEDs catalog due to one of the above reasons. The right panel of \autoref{badfit} demonstrate a perfect example on the importance to the inclusion of NIR data in the SED analysis for proper estimation of parameters.

Therefore, we finally catalog 347 candidate WDMS binaries with well-fitted SEDs. 
The catalog of 347 candidates is available on \href{https://zenodo.org/records/19270085}{Zenodo} and at the CDS with the following information: \gaia\ DR3 and GALEX IDs, stellar parameters, and the values of $\chired$, Vgf$_b$, and $\lvert f_{residue} \lvert$ in the NUV and the next bluest optical bands. A representative part of the catalog is presented in \autoref{tab:source_table}.
We also provide the catalog of 1,070 badly-fitted candidates on \href{https://zenodo.org/records/19270085}{Zenodo} and at the CDS containing their positions, \gaia\ DR3 and GALEX IDs. 
SEDs of all the well-fitted WDMS binary candidates as well as badly fitted candidates are also available on \href{https://zenodo.org/records/19270085}{Zenodo} and at the CDS. We  also provide a separate catalog of all the 6,559 UV bright sources, identified based on shift in the CMDs (blue points in the \autoref{selection}), which includes coordinates, IDs and magnitude of all photometric surveys used in this study.


\begin{figure}
	\includegraphics[width=\columnwidth]{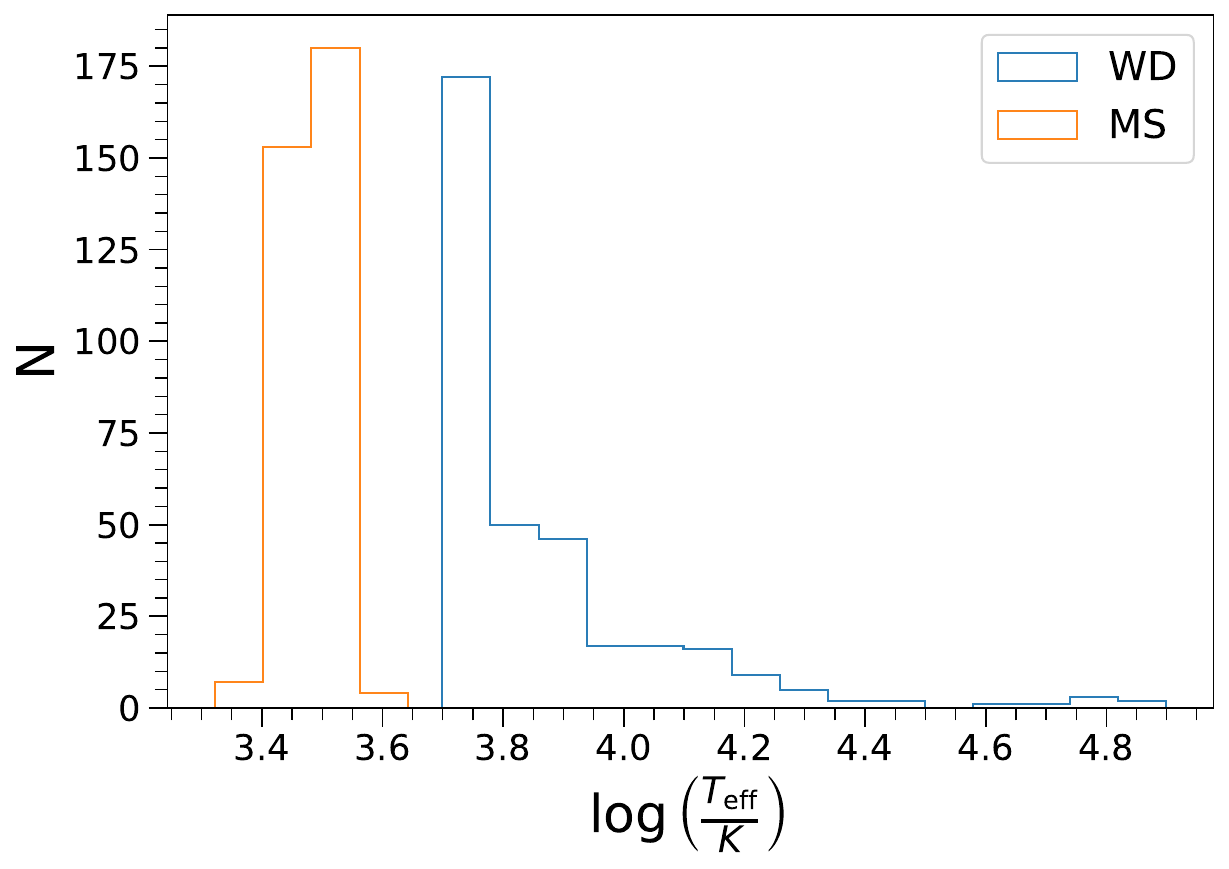}
    \caption{Distribution of $\teff$ for the WDs (blue) and MS (orange) stars in our candidate WDMS binaries. The median, 25th, and 75th percentiles for the MS s (WD) is $\teff/\rm{K}=3100^{+100}_{-200}$ ($6250^{+2000}_{-750}$). 
    }
    \label{teff}
\end{figure}

\begin{figure}
	\includegraphics[width=\columnwidth]{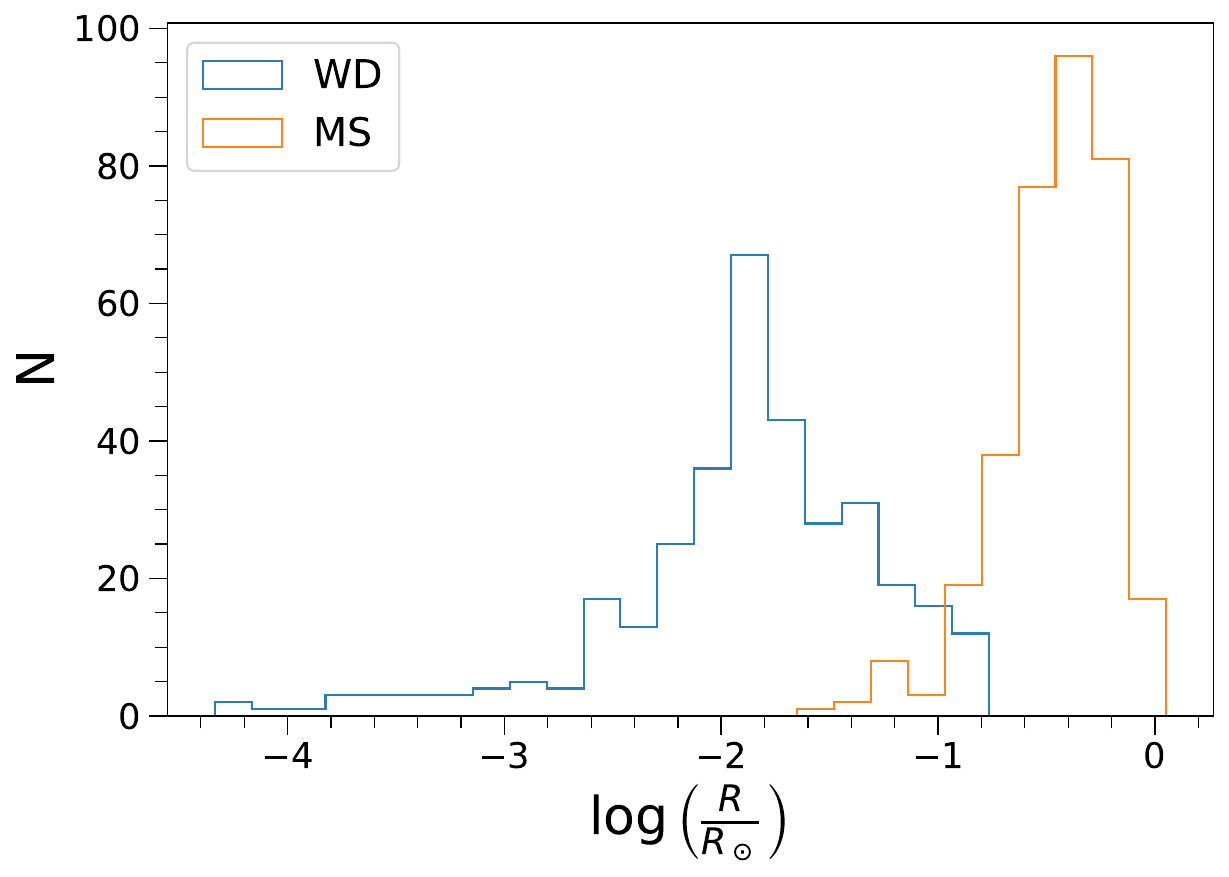}
    \caption{Same as \autoref{teff}, but showing the stellar radii for WDs (blue) and MS stars (orange). The median, 25th, and 75th percentiles for the MS stars (WDs) is $\log(R/R_\odot)=-0.40^{+0.14}_{-0.18}$ ($-1.82^{+0.39}_{-0.28}$). 
    }
    \label{radius}
\end{figure}


\section{Results and Discussion} \label{discussion}

We could fit the observed SEDs using composite model spectra of WDMS binaries for 347 sources and are marked with orange plus signs in \autoref{selection}.  
WDMS binary fits using the VOSA tool provide us with best-fit stellar parameters for both companions. 
In particular, $\teff$ and radius estimates from SEDs are robust \citep{bayo2008}. All stellar properties of each source are summarized in \autoref{tab:source_table}. We show the distributions of $\teff$ and radii for the MS stars (orange) and the WDs (blue) in \autoref{teff} and \autoref{radius}. We find that in our sample, the MS (WD) stars have $\teff/\rm{K}=3100^{+100}_{-200}$ ($6250^{+2000}_{-750}$). The corresponding numbers for the radii of the MS (WD) stars are $\log(R/R_\odot)=-0.40^{+0.14}_{-0.18}$ ($-1.82^{+0.39}_{-0.28}$).\footnote{The numbers in all cases denote the median. The lower and upper errors denote the 25th and 75th percentiles, respectively.}  
We find that most of the MS components in our catalog of WDMS binaries are M spectral types based on the $\teff$ vs spectral type relation \citep{pecaut2013}.

\subsection{WD mass}
\label{S:wd_mass}
The SED fitting process is found to be less sensitive to $\log\ g$\footnote{check the section 5.1 in VOSA help page https://svo2.cab.inta-csic.es/theory/vosa/helpw4.php?otype=star\&what=intro\#}, however, mass estimation is dependent on $\log\ g$. Therefore, a small change in the $\log\ g$ value can introduce a significant variation in mass estimations. Therefore, instead of directly using the $\log\ g$ values obtained from VOSA, we consider the WD evolutionary models created by \citet{Bedard2020} to calculate the masses of WDs. We assume that the WDs in our sample have CO cores with Hydrogen atmospheres. We use $\teff$ and bolometric luminosity ($\Lwd$) of the WDs obtained from our SED fittings as inputs to the WD evolutionary models to estimate their masses and cooling ages.

\begin{figure}
	\includegraphics[width=1.0\columnwidth]{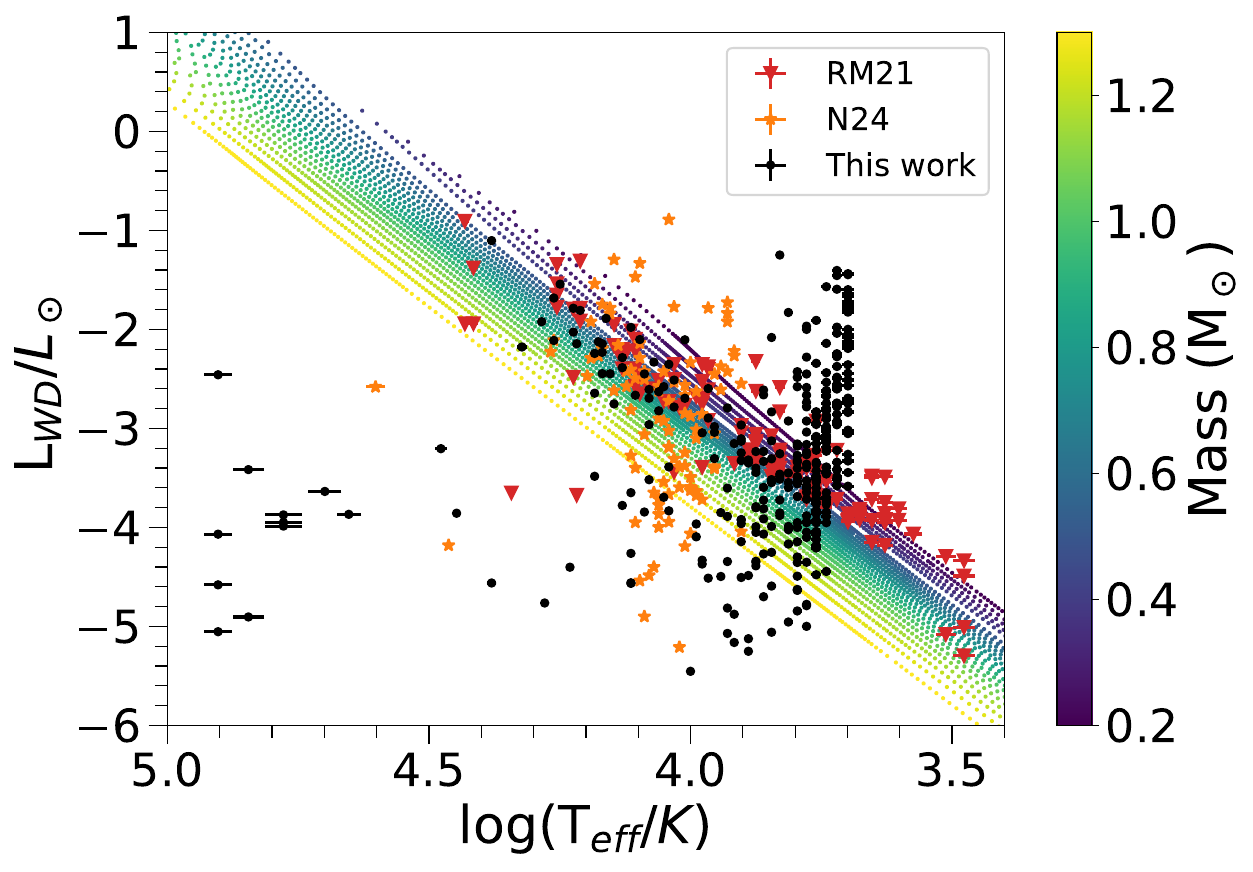} 
    \caption{$\teff$ vs $\Lwd$. Model cooling sequences for CO WDs with a Hydrogen atmosphere are shown where the color code denotes WD mass \citet{Bedard2020}. The black, orange and red points are the WDs identified in this work, \citetalias[][]{nayak_2024_wdms} and in 
    \citetalias[][]{Rebassa2021}, respectively.   
    }
    \label{wd_mass}
\end{figure}

In \autoref{wd_mass}, we show $\Lwd$ as a function of $\teff$ for the WDs in our sample with black dots. WD cooling tracks from models \citep[][]{Bedard2020} are overlaid. The cooling tracks are color-coded with WD mass. In general, for the same cooling age, lower-mass WDs would have lower $\teff$. On the other hand, over time, the WDs would cool along the sequences shown in the figure for any particular mass. 
We estimate the masses of the WDs from the model cooling tracks using linear interpolation.

We identify 121 candidate WDs in our sample with luminosities ($\Lwd$) exceeding the model evolutionary tracks. These WDs are likely to have lower masses than the $0.2\,M_\odot$ limit in CO-core WD models by \citet{Bedard2020}. Additionally, we find 56 candidate WDs with $\Lwd$ below the model evolutionary tracks. Consequently, we were unable to estimate the masses of these WD candidates, 
as we chose not to extrapolate beyond the model limits.

We could estimate masses for 170 WDs and they are distributed 
in the range $\sim$0.2 to 1.3 $M_\odot$ (\autoref{wd_mass}). 
The individual masses of these candidate WDs are listed in \autoref{tab:source_table}. Out of 170 candidates, we identify 16 ($\sim$9\%) ELM-WDs with masses $\le0.3\, M_\odot$. 
Detecting ELM-WDs is extremely challenging due to their low luminosity and low-mass companions. \citetalias{Rebassa2021} reported 35 ($\sim$31\%) ELM-WDs and \citetalias{nayak_2024_wdms} reported 5 ($\sim$ 8\%) ELM-WDs with MS companions in 100 pc volume. 
The fraction of ELM-WDs identified in this work ($\sim$9\%) is consistent with the value reported by \citetalias{nayak_2024_wdms} ($\sim$8\%), but significantly lower than the $\sim$31\% reported by \citetalias{Rebassa2021}. This discrepancy likely arises from differences in sample selection. \citetalias{Rebassa2021} focused exclusively on sources located in the gap region between the WD and MS in the optical CMD.  
In contrast, both this study and \citetalias{nayak_2024_wdms} include the sources which are majorly located on the MS with a small fraction on the gap region. 
Hence, differences in sensitivity to cooler WDs might have contributed to the variation in reported fractions.

Out of 16 ELM-WDs, only one (759601941671398272) was previously classified as DC+M4.5Ve WDMS binary based on SIMBAD. Nine out of 16 ELM-WDs are previously identified by \citet{Jackim_2024_WDMS_galex, Li_2025_30k_wdms}. Seven ELM-WD MS binaries are newly identified in this study. We also found a single epoch spectroscopic observation of one source (2246676707976104960) in the SDSS-V survey. The spectrum shows H-Balmer emission lines with no prominent excess flux in the bluer part of the spectrum. Previous spectroscopic surveys such as the SDSS magnitude-limited sample \citep[][]{Rebassa_2010_wd_sdssVII, Rebassa2012a, Rebassa2013a} cataloged only a few ELM-WDs with MS stars as their companion. A recent study by \citet{Jackim_2024_WDMS_galex} cataloged more than 900 potential ELM-WDs within binary systems. 
Though many ELM-WD+MS candidates are recently being detected \citep[][]{Jadhav_2019, Jadhav_2023, Subramaniam_2020, Khurana_2023, Sidharth_2024_WD_towards_SMC}{}{}, their formation channels remain poorly understood. Further detailed spectroscopic follow-up and population studies are necessary to gain better insights into their formation processes.

\subsection{Contamination from active M-dwarf stars}\label{sec:active_M-dwarf}

As shown in \autoref{wd_mass}, the WDs in several WDMS candidates lie outside the WD cooling sequence and their MS companions mostly belong to the M-dwarf populations. Since M-dwarf stars are known to have chromospheric activities which cause the presence of excess emission mainly in the UV regions \citep[][]{ Martinez_2010_active_MS, Boro_Saikia_2018_active_MS, Gomes_2021_active_MS}. The reconfiguration of strong magnetic fields in the outer atmosphere of these low-mass M-dwarf stars can also lead to the release of large amounts of energy in X-rays and UV in the form of eruptions or flares \citep{Vallee_2003, Samus_2017}. If the \galex\ observations are obtained during the flaring event, we will get the wrong estimations of the sources' magnitudes. 
Therefore, active M-dwarf stars can mimic WDMS binary candidates in the NUV-optical CMD. There have been extensive studies dedicated to understanding the phenomenon of stellar flares and chromospheric activities using X-ray, UV and optical observations \citep{ Martinez_2010_active_MS, Boro_Saikia_2018_active_MS, Gomes_2021_active_MS, Newton_2017_rotation_flares,  Brasseur_2019_flares, Rekhi_2023_flares, Li_2024_flares}. We compare our catalog with literature studies to find if we misclassified any active M-dwarf as WDMS binary.


\begin{figure}
	\includegraphics[width=1.0\columnwidth]{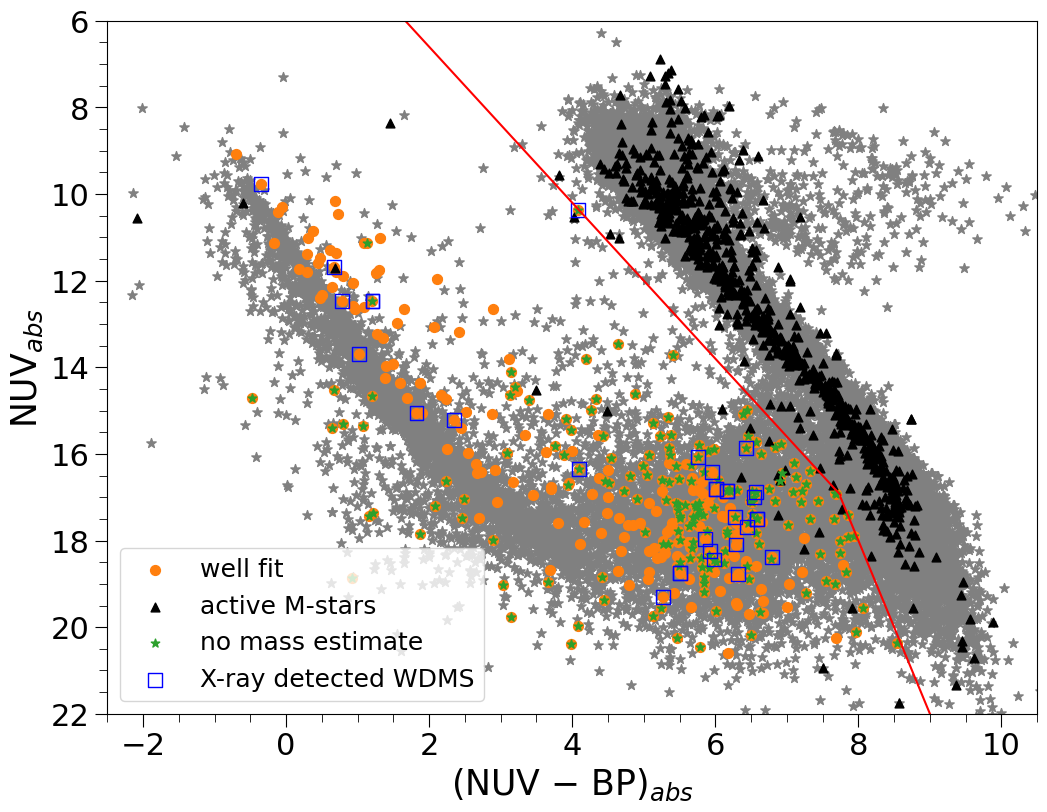}
    \caption{ Compare the location of WDMS binary candidates with active MS stars and X-ray detected WDMS candidates in the NUV-optical CMD. }
    \label{compare_active_MS}
\end{figure}


\citet{Boro_Saikia_2018_active_MS} studied the chromospheric activity of 4454 cool stars, combining archival HARPS spectra and multiple other surveys.  
However, there are no crossmatch with our catalog. 
We do not find any match with the catalog by \cite{Gomes_2021_active_MS}, where the authors studied the chromospheric activity of 1674 FGK stars using the HARPS archival data. We also do not find any match with \cite{Martinez_2010_active_MS}, which presented spectroscopic analysis of chromospheric activity for nearby (d$<$25pc) cool (spectral types F to K) stars. 
\cite{Newton_2017_rotation_flares} presented an optical spectroscopic study of 2202 low mass (between 0.1 to 0.6 M$_\odot$) stars in the solar neighborhood to study their stellar flares activity and only one source is in common with our catalog. 
We find no match with \cite{Brasseur_2019_flares}, where the authors detected short duration ($<$ 5 minutes) NUV flares of $\sim$1000 stars using \galex\ data. We do not find any crossmatch with \cite{Yang_2017_flares_kepler_field}, where the authors studied the flaring activity of 540 M-dwarfs in the Kepler field. 
\cite{Medina_2020_flares_15pc} presented a spectroscopic study of stellar flares in 125 low mass stars (between 0.1 to 0.3 M$_\odot$) within 15 pc volume, we do not find any common sources with our catalog. 
We find one match with \cite{Li_2024_flares} and no match with \cite{Rekhi_2023_flares} where the authors studied on flares and chromospheric activity of M-dwarf using \galex\ data \citep{Rekhi_2023_flares, Li_2024_flares}.  
Therefore, we find 2 sources in our catalog which are previously cataloged as active MS stars.

Active MS stars are also expected to have X-ray counterparts. We check for their X-ray counterparts in various catalogs and found 7 matches in Chandra \citep[][]{Evans_2010_CHANDRA}{}{}, 13 matches in XMM-Newton \citep[][]{Saxton_2008_XMM-Newton-slew, Webb_2020_XMM-Newton-epic}{}{}, and 12 matches in ROSAT all-sky survey data \citep[][]{Boller_2016_2RXS}. In total, we get X-ray counterparts for 27 sources in our catalog. 
In \autoref{compare_active_MS}, we overlaid the active MS stars located within 100 pc from the catalogs by \citet{Boro_Saikia_2018_active_MS, Newton_2017_rotation_flares, Gomes_2021_active_MS} and those 27 WDMS candidates with X-ray counterparts on NUV $-$ optical CMD to investigate their location in the CMD and assess the potential contamination of these stars in our catalog. We also highlighted the WDMS candidates where we could not estimate the masses of WDs (blue open boxes). The figure shows that a large fraction of WDs with no mass estimations are located in the $5.5 \lessapprox (NUV - BP) \lessapprox 8$ color range, close to the MS branch. Majority of WDMS candidates with X-ray counterparts are also located in this color range. 
While active MS stars are mainly located on the MS branch in NUV$-$optical CMDs as well with a few in the above color range. 
If we impose a cut-off on (NUV$-$optical) color and consider only the sources hotter than (NUV$-$BP)$_{abs}$ = 5.5, the number of binary candidates with well-fitted SEDs decreases to $\sim$54\% than the original catalog (187 sources). The fraction of WDs without mass estimates decreases to $\sim$40\% (74 out of 187), compared to $\sim$51\% (177 out of 347) in the full sample. Similarly, the fraction of sources with X-ray counterparts decreases to $\sim$6\% (11 out of 187) relative to $\sim$8\% (27 out of 347) in the original catalog. We find that adopting a stricter (NUV$-$optical) color cut does not lead to a statistically significant reduction in contamination. 
Therefore, we conclude that a small fraction of WDMS candidates in our catalog might be contaminated by active MS stars. 

Spectroscopic observations are required to better distinguish WDMS binaries from the active MS stars. We cross-match these 27 X-ray detected sources with LAMOST DR11 catalog and found 12 in common. \gaia\ DR3 2560009007603950720 is categorized as DC+M WDMS binaries in the catalog. 10 out of the rest 11 sources exhibit strong H-Balmer emission lines ($H\beta$, $H\gamma$, $H\delta$, $H\epsilon$). Five of them appears to have variation in the emission strength in multi-epoch observations, suggesting that the emissions could be coming from close or interacting binaries. However, these emissions can also come from the active M-dwarf due to flaring activity. A more dedicated spectroscopic study to quantify the strength, presence UV  excess and period of variability is necessary.  Follow-up deep FUV photometric observations will also be beneficial in separating these two populations \citep{Anguiano_2022}. However, it is important to note that the presence of stellar activity in an M-dwarf does not directly rule out the possibility of having a WDMS binary, as post-common envelop systems with active M-stars are reported in the literature \citep{Muirhead_2013_post_CE_Mdwarf}.

\begin{figure}
	\includegraphics[width=1.0\columnwidth]{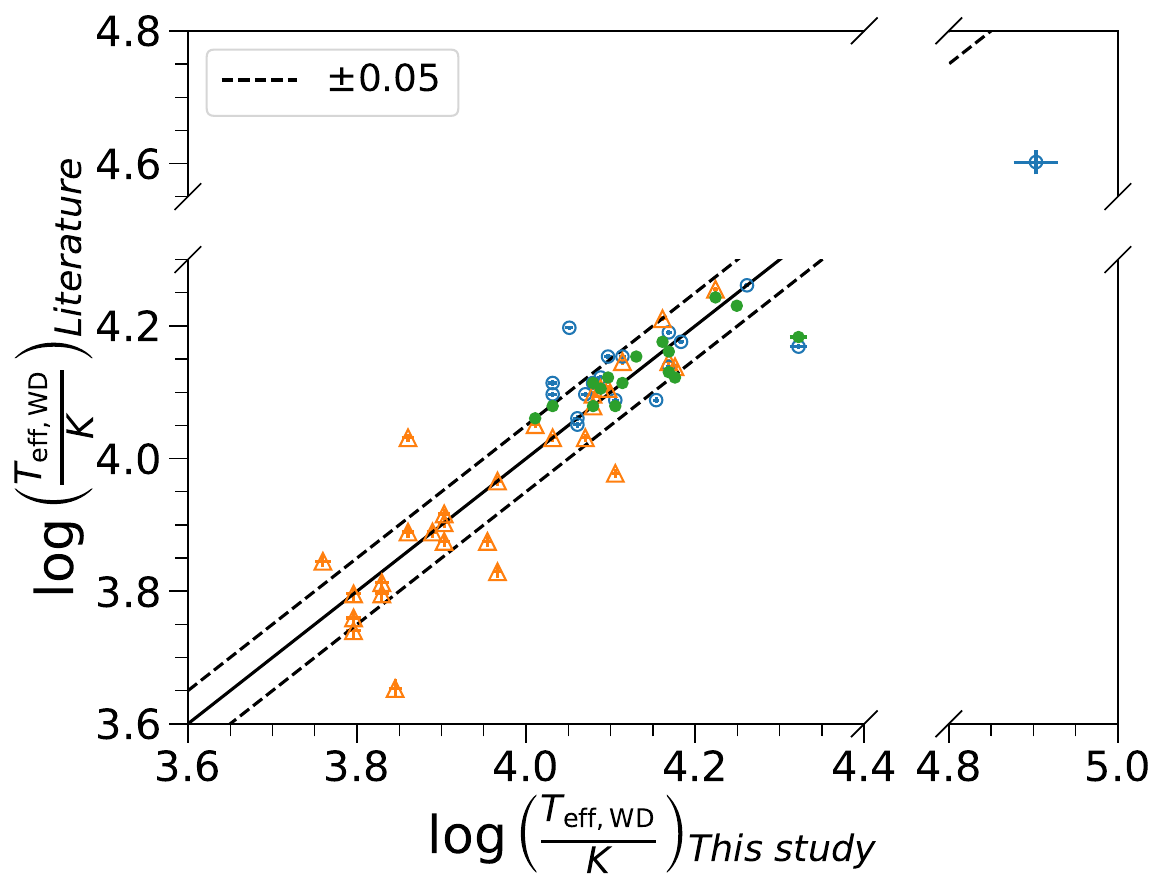} \includegraphics[width=1.0\columnwidth]{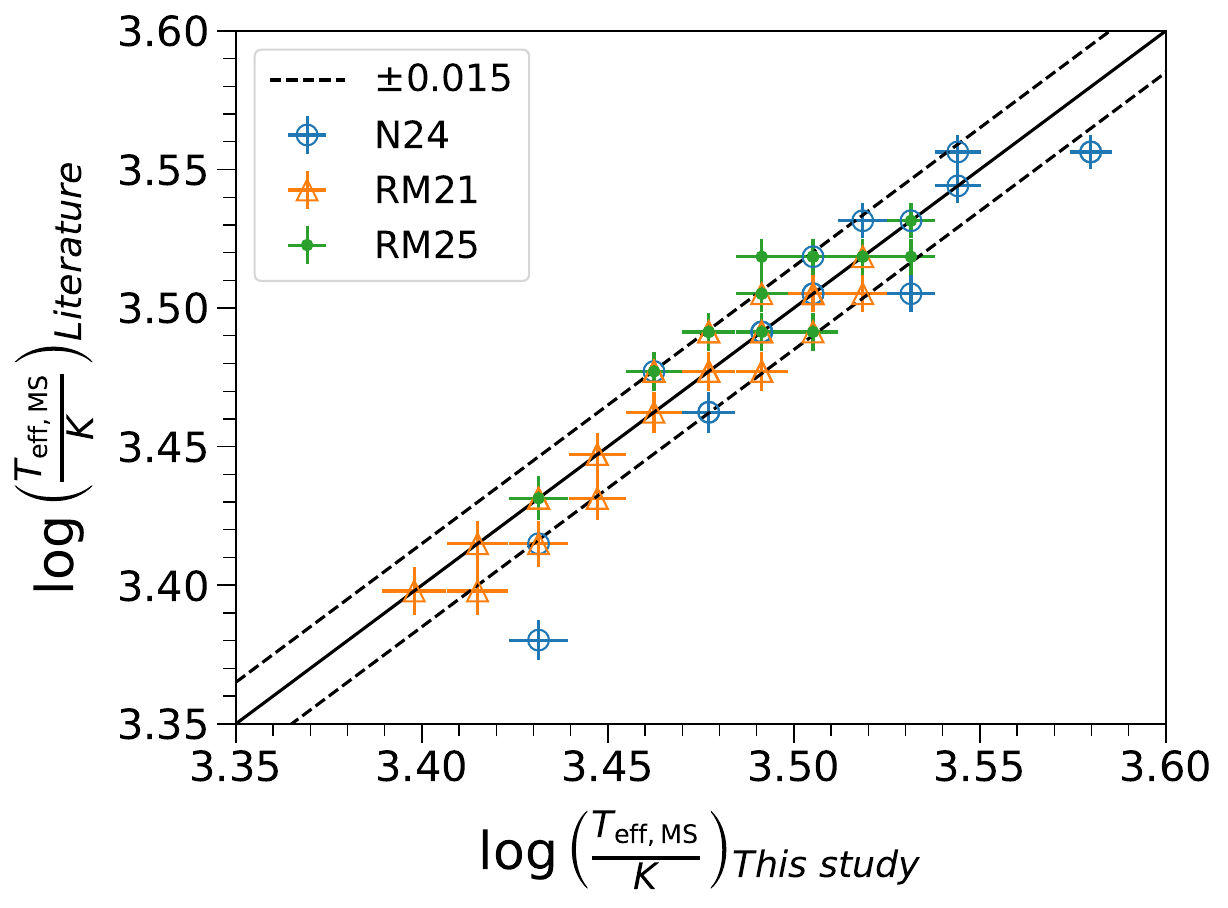} \includegraphics[width=1.0\columnwidth]{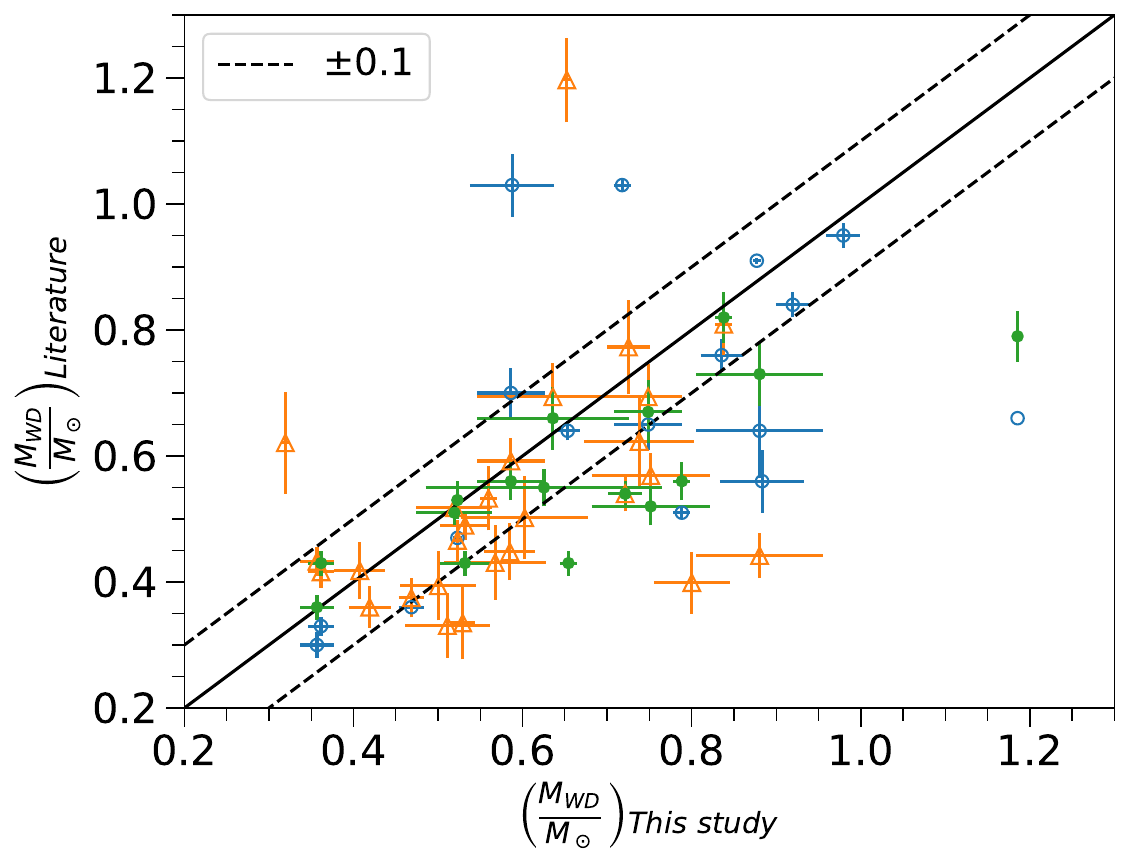}
    \caption{ Comparison of WD temperature, MS temperature and WD mass with the literature estimations. The solid lines in all the panels represent 1:1 relation, while the dashed lines are deviation boundaries to our estimates. }
    \label{compare_parameters}
\end{figure}

\begin{figure}
	\includegraphics[width=1.0\columnwidth]{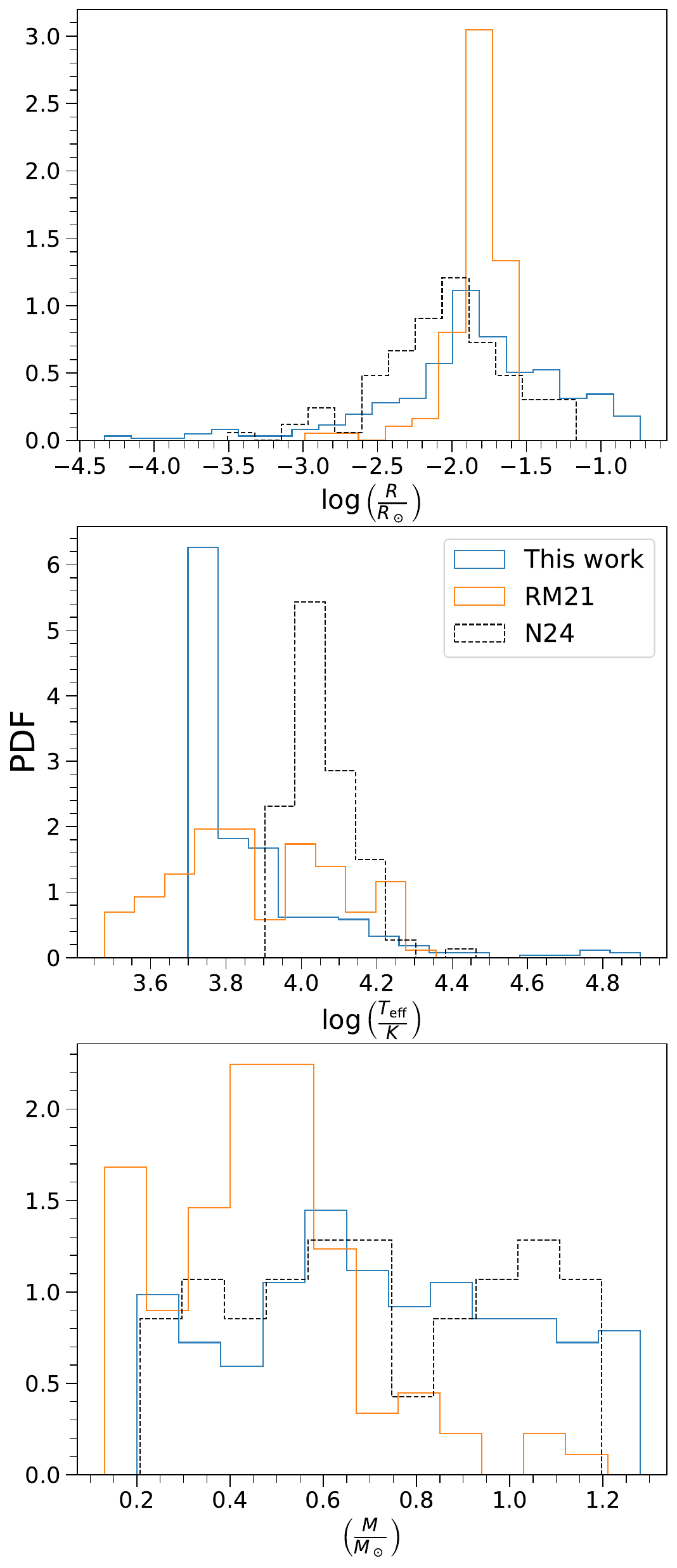}
    \caption{ Comparison of properties for WDs identified in this work (blue), \citetalias{Rebassa2021} (orange) and \citetalias{nayak_2024_wdms} (black). From top to bottom, we show distributions of radius, $\teff$, and mass, respectively. }
    \label{compare}
\end{figure}


\subsection{Compare with previous WDMS catalogs} \label{comparison}

The recent WDMS catalogs focused on a volume-limited sample of 100 pc within the solar neighborhood are provided by \citetalias{Rebassa2021} and \citetalias{nayak_2024_wdms}. 
\citetalias{Rebassa_2025_mag_limited} recently extended the study of \citetalias{Rebassa2021} to a few kpc to generate a magnitude-limited catalog of 1,312 WDMS binaries having Gaia XP spectra and provided stellar parameters for 435. Out of 1,312 sources, 90 are within 100 pc with stellar parameters provided for 19 sources.  These studies also performed binary SED analyses to derive stellar parameters. While \citetalias{nayak_2024_wdms} focused on sources with both FUV and NUV observations, \citetalias{Rebassa2021} and \citetalias{Rebassa_2025_mag_limited} also included sources with only NUV or without UV data. A comparison between our results and those from these works provides an additional check on the reliability of our estimates and methodology. We find 18 crossmatch with \citetalias{nayak_2024_wdms}, 29 with \citetalias{Rebassa2021}, and 31 (but parameters estimated for 16) with \citetalias{Rebassa_2025_mag_limited}. 
In \autoref{compare_parameters}, we compared WD temperature (top panel), MS temperature (middle panel) and WD mass (bottom panel) of these common sources. The solid lines in all the panels correspond to 1:1 relation, while the dashed lines represent deviation boundaries to our estimates: $\pm$0.05 in log scale for WD temperature, $\pm$0.015 in log scale for MS temperature, and $\pm$0.1 $M_\odot$ for WD mass. We found that both WD and MS temperature estimations are in good agreement with literature estimates within the deviation boundaries. When we inspect SEDs of the sources having relatively larger deviation with \citetalias{Rebassa2021} in lower temperature regime, we noticed that observed SEDs are badly fitted in \citetalias{Rebassa2021} and do not satisfy the goodness criteria defined by us as well as by \citetalias{nayak_2024_wdms}\footnote{check out the SEDs of these \gaia\ DR3 IDs (2155188926705745536, 2630815357409558400, 3668430081179262208, 4384149753578863744, 6182278665776280320,  in http://svocats.cab.inta-csic.es/wdw4/index.php?action=search}. 
Our SED fittings provided better results than \citetalias{Rebassa2021} for these sources, which could be due to the recent upgrade to fitting method in the VOSA tool, namely, detection of outliers among the photometric data before the SED fitting and improved calculation of the total flux in the chi-square fitting (details can be found on the \href{https://svo2.cab.inta-csic.es/theory/vosa/helpw4.php?otype=star\&action=help\&what=news}{VOSA website}). 

In the higher temperature regime, we have one source common with \citetalias{nayak_2024_wdms} that shows a significant deviation, with our analysis overestimating the WD temperature. This discrepancy highlights the importance of FUV observations for accurately constraining very hot WDs ($T_{\rm eff} \gtrsim 20{,}000$ K; $\log T \gtrsim 4.3$), whose spectral energy distributions peak at shorter wavelengths than NUV band. As mentioned earlier in the section 2.3, our selection method is intrinsically biased toward detecting the WDMS binaries with cooler WDs, due to relying solely on NUV band in the UV regime, which is primarily sensitive to temperatures of $\sim$10,000–16,000 K. As a result, we noticed that temperature estimates for hotter WDs are less well constrained and tend to be systematically overestimated. We identified 16 WDMS binary candidates in this regime whose WD parameters should be treated with caution and would benefit from follow-up FUV observations, which are sensitive to hotter WD temperatures ($\lesssim$40,000 K).   
As the deviation in the estimation of WD temperature also reflected in the estimation of WD mass, producing systematically higher masses with an average offset of $\sim$0.1 $M_\odot$, while temperatures of MS companions remain comparatively less affected. We infer that followup observations with upcoming UV missions like INdian Spectroscopic and Imaging Space Telescope \citep[INSIST;][]{purni_2022_INSIST, Sriram_2023_INSIST}, the Ultraviolet Explorer \citep[UVEX;][]{Kulkarni_2021_UVEX}, and The Cosmological Advanced Survey Telescope for Optical and UV Research \citep[CASTOR;][]{Cote_2019_CASTOR}, which offer coverage in both FUV and NUV, will significantly improve in constraining the WD parameters. In addition, their deeper photometric limits (2-4 magnitudes beyond \galex), in particular the UVEX in wide-field survey mode, will help us in detecting even fainter WDMS systems and improve sample completeness in the solar neighborhood.

In \autoref{compare}, we compare distribution of our estimated mass, radius, and $\teff$ of the WDs with those found in \citetalias{Rebassa2021} (orange) and \citetalias{nayak_2024_wdms} (black). The figure shows that  
peak in $\teff$ distribution in our study is found to cooler than \citetalias{nayak_2024_wdms} while peak in the radii distribution is found to be similar to \citetalias{Rebassa2021} and \citetalias{nayak_2024_wdms}. 
The figure also shows that our detected WDs have similar mass distribution to that found in \citetalias{nayak_2024_wdms} but more massive than those detected in \citetalias{Rebassa2021}.

Recently, \citet[][]{Jackim_2024_WDMS_galex} provided the most extensive catalog of single and binary WD using \gaia\ and \galex\ data, where the authors selected the WD from UV and NUV-optical CMDs and search for their companions in the non-WD regions on optical CMD. We find 154 common sources with their catalog, out of which 9 were reported as single WD. However, out of these 9, two sources \gaia\ DR3 2560009007603950720 and \gaia\ DR3830644063706298496 are categorized as DC+M and DA+M WDMS binaries, respectively, while \gaia\ DR3 704748371716051072 shows signature of WDMS binaries in one of the two epoch of observations in the LAMOST DR11 optical spectroscopic catalog (see \autoref{sec:spectroscopy}). The SED corresponds to \gaia\ DR3 ID 2560009007603950720 is also shown in the right panel of \autoref{sed}, which clearly shows the presence NIR excess which highlights the significance of multi-wavelength studies in detecting and characterizing WDMS binaries. 
WD $\teff$ for the common sources are found to be a good match following a 1:1 relation except for a few candidates, while WDs in our catalog appear relatively more massive than that estimated by \citet[][]{Jackim_2024_WDMS_galex}. 
We notice that the use of higher extinction values and not inclusion of the NIR data points could be the reasons for the deviation in the estimated parameters and for defining the two binaries as single WDs. Since sources in this work are located within 100 pc, extinction is expected to be negligible which is also found in the MWDUST 3D extinction map solution. However, \citet[][]{Jackim_2024_WDMS_galex} used a 2D extinction map from \cite{Schlegel1998}, which provides significantly larger extinctions for these solar neighborhood sources. As the UV region is more sensitive to extinction, the use of higher extinction values will make the intrinsic luminosity of the WD brighter and less massive on the WD cooling sequence track (see Fig. \ref{wd_mass}).

We do not find any match with the large surveys on WDMS binaries by \citet[White Dwarf Binary Pathways Survey;][]{WDS5_Ren2020}. Their sample primarily focuses on the selection of WDMS binaries with early type MS companions (i.e., AFGK type, with $T_{\rm eff} > 4000$ K), whereas the selection method adopted in this work prefer to identify WDMS binaries with later type companions (i.e., M dwarfs).

We find that 9 WDMS candidates in our catalog match with the catalog of WDMS binaries detected using LAMOST \citep{Ren2018_wdms_LAMOST_DR5} and 10 matches with \cite{Anguiano_2022}. 
There are 4 WDMS binaries from our catalog already identified by the SDSS spectroscopic catalog of WDMS binaries \citep[][]{Rebassa_2010_wd_sdssVII} and 9 are previously cataloged in SDSS photometric catalog of WDMS binaries \citep[][]{Rebassa2013b}. 
We find three matches with the catalog provided by \cite{Xabier_2025_ML_WDMS}, where the authors applied unsupervised machine learning algorithm to Gaia XP spectra and identified 993 WDMS binary candidates.
\cite{Li_2025_30k_wdms} recently studied approximately 30,000 WDMS binary candidates using Gaia XP spectra and 1700 of them are high-confidence systems confirmed through spectral fitting. We find 20 sources common with their catalog. 
We also cross-matched our catalog with that of single WD based on \gaia\ -EDR3 by \cite{Gentile_2021_Gaia_WD} but did not find any match.  
Comparison with \gaia\ DR3 non-single star (NSS) catalog \citep[][]{Gaia_DR3_NSS} provides 2 match with ours. 
The match with the NSS catalog also highlights the importance of multi-wavelength study from UV to IR to constrain the binary properties for sources in the NSS catalog \citep{Ganguly_2023}.  
The overlap between our catalog and previously identified candidates confirm the validity of these candidates and supports the effectiveness of our method for identifying new WDMS candidates.
In this study, we have identified and cataloged 347 WDMS binary candidates and 188 of them are newly detected and are not classified as WDMS binaries or active M-stars in the literature.


\subsection{Compare with Spectroscopic Survey and SIMBAD}\label{sec:spectroscopy}

\begin{figure}
	\includegraphics[width=1.0\columnwidth]{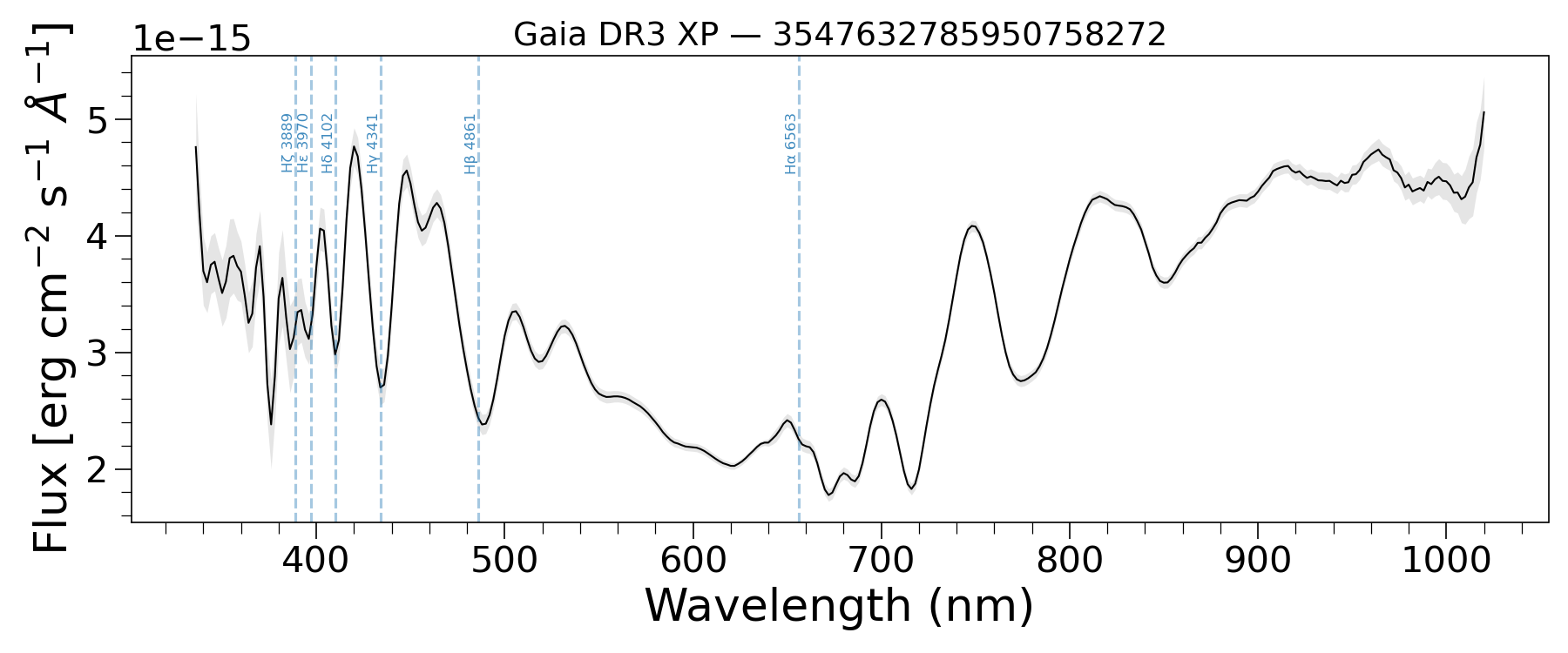}
    \includegraphics[width=1.0\columnwidth]{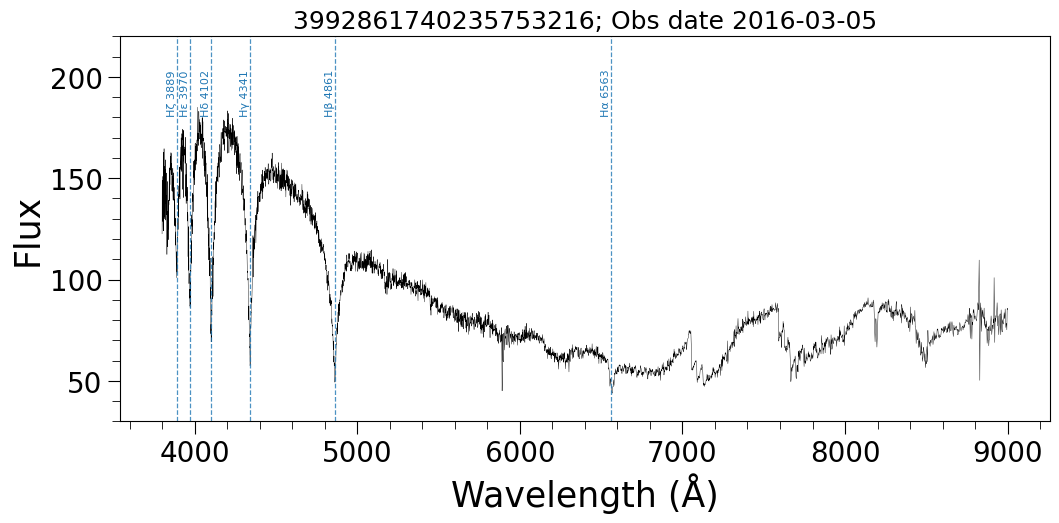}
    \includegraphics[width=1.0\columnwidth]{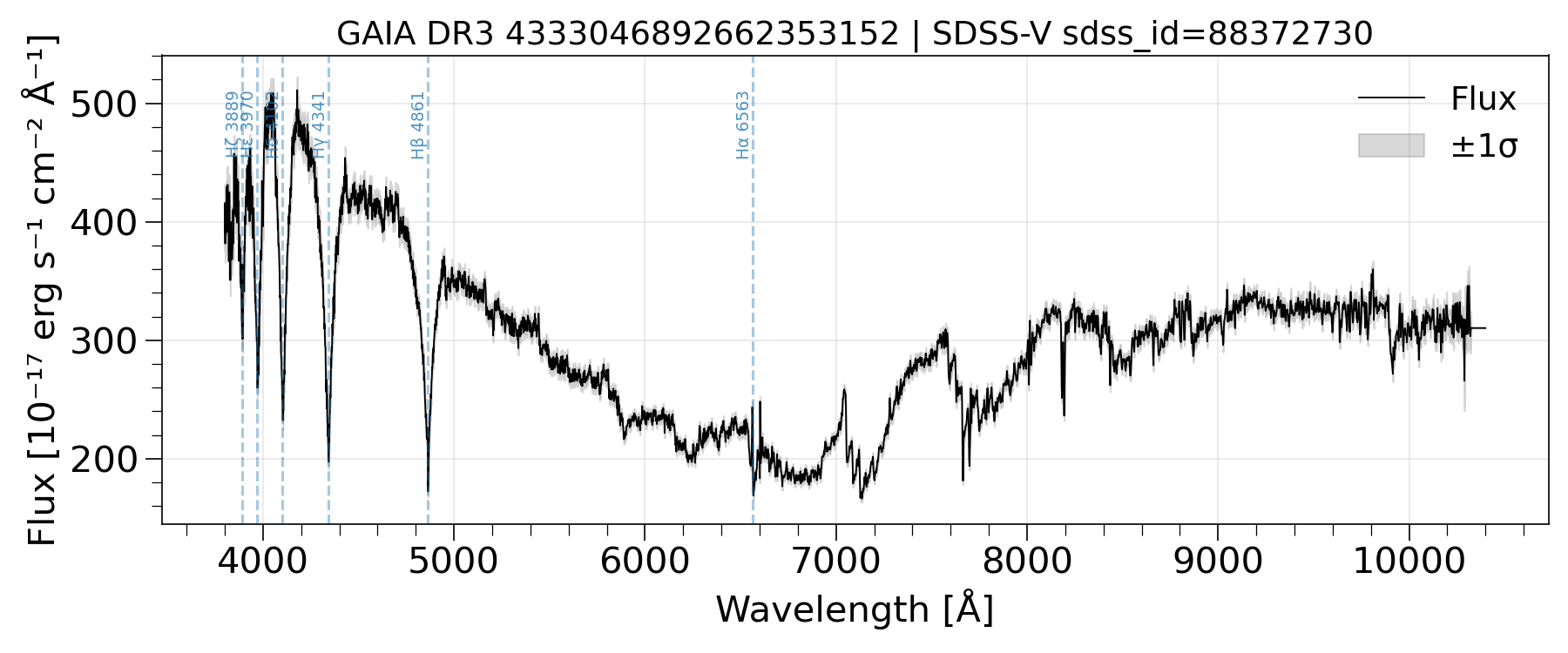}
    \caption{ (Top panel:) An example of low resolution \gaia\ XP spectrum of \gaia\ DR3 3547632785950758272.  
    (Middle panel:)  An example LAMOST spectrum of Gaia source id 3992861740235753216. This source is defined as DAZ+M WDMS binary in the LAMOST catalog. (Lower panel:) An example SDSS spectrum of \gaia\ DR3 947545965334561280. In all the panels, error in flux are represented as gray shaded region and H-Balmer lines are marked in blue. All the three spectra show the presence of strong H-Balmer emission lines and excess emission towards the bluer part of the spectrum, representing the signatures of a WD in binary with a MS star. }
    \label{spectra}
\end{figure}

\gaia\ DR3 also provided low resolution BP/RP (XP) spectra of more than 200 million stars. We found that 208 sources have \gaia\ XP spectra, out of 347. In the spectra, we expect to see excess emission towards bluer part of the optical spectra and/or presence of H-Balmer absorption lines ($H\beta$, $H\gamma$, $H\delta$, $H\epsilon$) \citep{Rebassa2016a}.  Following the classification scheme introduced by \citetalias{Rebassa_2025_mag_limited} (see their Fig.2), we identify 23 sources showing the signatures of WDMS binary, as shown in the top panel of \autoref{spectra}. In the spectrum, H-Balmer absorption lines are also marked. A complete crossmatch catalog of these 208 sources along with their Gaia XP spectra are available on \href{https://zenodo.org/records/19270085}{Zenodo}. 

We further compare our catalog with available spectroscopic surveys to find out the signatures of WDs. We found 93 match with optical spectroscopic survey LAMOST DR11\footnote{https://www.lamost.org/dr11/v2.0/}. 16 of them show prominent detection of H-Balmer absorption lines and/or excess emission in the bluer part of the spectra, suggesting them to be confirmed WDMS binaries. An example spectrum is shown in the middle panel of \autoref{spectra}. Out of these 16 sources, five are classified as DA+M WDMS binaries (\gaia\ IDs: 3992861740235753216, 4437836226304032384, 704748371716051072, 830644063706298496, 881086019353249280) in the LAMOST catalog and one as DC+M WDMS binary (2560009007603950720). \gaia\ DR3 704748371716051072 shows signatures of WDMS binary in one of the two epoch of observations, indicating eclipsing nature of this source. Two sources are classified as double stars (\gaia\ IDs: 1006621281985546240, 1911816151065187456). 
Spectral classification of majority of the LAMOST detected sources are M-type, which matches with our finding using SED analyses. \gaia\ DR3 IDs, $\teff$ for WD and MS from SED analyses, and combined $\teff$ values and spectral type classification from LAMOST catalog are listed in \autoref{tab:lamost_match}. A complete crossmatch catalog of 93 sources along with LAMOST spectra are available on \href{https://zenodo.org/records/19270085}{Zenodo}.

We found 22 match with optical spectroscopy survey SDSS DR19\footnote{https://skyserver.sdss.org/dr19, https://dr19.sdss.org/zora/}. Nine of them show prominent H-Balmer absorption lines and presence of excess flux in the bluer part of the spectrum, suggesting them to be confirmed WDMS binaries. An example spectra of \gaia\ DR3 947545965334561280 is presented in the bottom panel of \autoref{spectra}. A dedicated spectroscopic study using all the available spectroscopic survey is necessary to examine the presence of small UV excess, estimation of temperature and $\log\ g$ parameters. A complete crossmatch catalog of 22 sources along with SDSS spectra are available on \href{https://zenodo.org/records/19270085}{Zenodo}.

We also matched our catalog with SIMBAD and found 274 common sources. Out of 274, 21 sources are classified as WDMS binaries, 7 are WDs and 62 are single MS stars of M spectral type. Out of 21 WDMS binaries, two are classified as spectroscopic binaries and one of them is cataclysmic variable (CV). MS companions of 19 are M spectral type, one is K spectral type and for one binary it is not defined. The spectral types of MS stars are in good match with the estimates from our method. \gaia\ DR3 IDs, $\teff$ for WD and MS from SED analyses, and SIMBAD source IDs and spectral type classification from SIMBAD catalog are listed in \autoref{tab:simbad_match}. Out of 7 common sources classified as WD, one (2843374388402149632) is mentioned as CV in SIMBAD, one (6461421956281723392) shows prominent binary signature in the \gaia\ XP spectra. A complete crossmatch catalog of 274 sources are available on \href{https://zenodo.org/records/19270085}{Zenodo}.

\subsection{Completeness of the Catalog}\label{completeness}

Through simulation \citetalias{Rebassa2021} showed that $\sim$91\% of total WDMS populations are hidden in the MS region of optical CMD. We cataloged and parameterized 347 WDMS binaries in this study, which are located either on the MS or gap region in optical \gaia\ CMD. Considering all the 6,559 UV bright sources are probable binary candidates, we are able to retrieve 5.3\% through multi-wavelength SED analyses. However, \galex\ survey covers only a fraction of the sky, mainly avoiding $\pm$10 degrees around the Galactic plane compared to all sky \gaia\ survey. We notice that $\sim$48\% of 100 pc \gaia\ data are located $\pm$10 degrees above the Galactic plane. Out of them, $\sim$83\% have reliable photometric or astrometric solutions as per the conditions mentioned in \autoref{data}. Therefore, considering the fractional sky coverage of \galex\ survey, completeness of our catalog has a lower limit of $\sim$2\%. 

The preferential coverage of intermediate and high Galactic latitudes by \galex\ implies that our catalog is biased toward lower-extinction environments and relatively older stellar populations. However, this spatial incompleteness is unlikely to significantly affect our primary demographic result, the prevalence of WDs with M-dwarf companions. The dominance of M-dwarf companions in WDMS binaries is mainly driven by the initial mass function and stellar evolutionary timescales, which favour long-lived low-mass secondary companions. 
Since our analysis is restricted to the 100 pc volume within the solar neighborhood, the stellar populations are well mixed and not dominated by very young populations. So, we do not expect the companion-type distribution to differ substantially within the Galactic plane.
Access to both FUV and NUV observations could potentially improve the completeness for WDMS binaries, however, the UV--optical selection method intrinsically favours WD systems with late-type (e.g., M-dwarf) companions due to contrast effects. Therefore, while GALEX sky coverage primarily impacts absolute completeness of the catalog, it is unlikely to significantly alter the inferred companion-type demographics within the surveyed footprint.

\smallskip
\smallskip

Our study demonstrates how \galex\ NUV and \gaia\ data can be used to identify those hidden WDMS populations and multi-wavelength SED analyses can be used to estimate their stellar parameters.
As our study used those sources which are located both on the MS and gap region of the optical CMD, and are detected majorly in \galex\ NUV, this study is complementary to many previous studies where both FUV and NUV data are used to identify WDMS binaries \citep{WDS5_Ren2020, Anguiano2020_wd_APOGEE, nayak_2024_wdms} or only the gap region in the optical CMD is targeted \citepalias[][]{Rebassa2021, Rebassa_2025_mag_limited}. We suggest a follow-up deep FUV observation of our catalog with Ultra-Violet Imaging Telescope (UVIT) on board AstroSat \citep[][]{tandon2017a, tandon2017b} which will be beneficial to better constrain the stellar parameters of WDMS binaries.  
However, a large number of candidates yet to be explored even within 100 pc, due to unavailability of deep UV observations or high-resolution spectroscopic observations in the Galactic plane. Given that obtaining  high-resolution spectroscopic observations for a large number of sources will be challenging, our study illustrates a simple pathway using multi-wavelength photometric data to identify WDMS binaries.  
We believe that implementing our method along with those demonstrated in the literature \citep{WDS5_Ren2020, Anguiano2020_wd_APOGEE, nayak_2024_wdms, Jackim_2024_WDMS_galex} on 
the upcoming UV missions like INdian Spectroscopic and Imaging Space Telescope \citep[INSIST;][]{purni_2022_INSIST}, the Ultraviolet Explorer \citep[UVEX;][]{Kulkarni_2021_UVEX}, will significantly advance the discovery of hidden WDMS binaries and contribute to completing a volume-limited WDMS sample. Once identified, a high-resolution spectroscopic follow-up study will help to improve their parameters as well as in better understanding their formation and evolutionary processes.


\section{Conclusions}
\label{conclude}
   \begin{enumerate}
      \item We have used NUV-optical CMDs as a tool to identify unresolved WDMS binaries within 100 pc. We have used \gaia\ DR3 for optical and \galex\  for NUV data to construct the NUV-optical CMDs.
      \item We have identified 347 WDMS binaries within 100 pc and 188 of them are newly reported.
      \item We have used a binary fitting algorithm available in the VOSA toolkit to determine $\teff$, bolometric luminosities, and radii of both the binary components. Our method predominantly identifies binaries with cooler WDs ( $<$10,000 K) compared to previous studies \citepalias[][]{Rebassa2021, nayak_2024_wdms, Rebassa_2025_mag_limited}. The MS components are found to be M spectral types. 
      \item We found a peak in the WD radii distribution at $\log(R/R_\odot)=$ $\sim -1.82$. 
      \item To determine the mass of WDs, we have used the WD evolutionary models by \citet{Bedard2020} and considered CO core and Hydrogen atmosphere. The WD masses range from $\sim$0.2 and 1.3 M$_\odot$. We have also identified 16 WDMS binaries with ELM-WD ($\le 0.3\,M_\odot$), which are likely formed due to mass loss from their progenitor stars. 
      \item This study presents a simple approach to identify and parameterize a large number of WDMS binaries combining multi-wavelength data. We suggest that implementing our method along with those demonstrated in the literature \citep{Ren2018_wdms_LAMOST_DR5, Anguiano_2022, Rebassa2013b, nayak_2024_wdms} on upcoming optical (\gaia\ DR4) and UV (INSIST, UVEX) surveys will lead towards the completeness of WDMS binaries in the solar neighborhood. 
   \end{enumerate}

\section{Data availability}

The catalogs of 347 WDMS binaries with well-fitted SEDs, 1,070 candidates with badly-fitted SEDs and 6,559 UV-bright sources are available on \href{https://zenodo.org/records/19270085}{Zenodo} and at the CDS. 
The SED figures for both well-fitted and badly-fitted SEDs are shared on \href{https://zenodo.org/records/19270085}{Zenodo} and at the CDS. 
All archival spectra retrieved for the 347 WDMS binaries from \gaia, LAMOST, and SDSS are made available on Zenodo, along with the cross-match catalogs. The catalogs hosted on Zenodo contain additional columns beyond those shared at the CDS (for example, all available columns from \textit{Gaia}, \galex, and other surveys), which may be useful for readers conducting follow-up studies.

\begin{acknowledgements}
      PKN acknowledges support from the Centro de Astrofisica y Tecnologias Afines (CATA) fellowship via grant Agencia Nacional de Investigacion y Desarrollo (ANID), BASAL FB210003. We thank the anonymous referee for an insightful and constructive review that significantly improved the clarity and quality of this work.
\end{acknowledgements}


   \bibliographystyle{aa} 
   \bibliography{WDMS_NUV_GAIA} 

\begin{thebibliography}{95}
\expandafter\ifx\csname natexlab\endcsname\relax\def\natexlab#1{#1}\fi

\bibitem[{{Andrews} {et~al.}(2022){Andrews}, {Taggart}, \&
  {Foley}}]{Andrews2022}
{Andrews}, J.~J., {Taggart}, K., \& {Foley}, R. 2022, arXiv e-prints,
  arXiv:2207.00680

\bibitem[{{Anguiano} {et~al.}(2020){Anguiano}, {Lewis}, {Corcoran},
  {Washington}, {Majewski}, {Allende Prieto}, {Mazzola}, {Badenes}, {Stassun},
  \& {Blondin}}]{Anguiano2020_wd_APOGEE}
{Anguiano}, B., {Lewis}, H.~M., {Corcoran}, K.~A., {et~al.} 2020, Research
  Notes of the American Astronomical Society, 4, 127

\bibitem[{{Anguiano} {et~al.}(2022){Anguiano}, {Majewski}, {Stassun},
  {Badenes}, {Daher}, {Dixon}, {Allende Prieto}, {Schneider}, {Price-Whelan},
  \& {Beaton}}]{Anguiano_2022}
{Anguiano}, B., {Majewski}, S.~R., {Stassun}, K.~G., {et~al.} 2022, \aj, 164,
  126

\bibitem[{{Bailer-Jones} {et~al.}(2021){Bailer-Jones}, {Rybizki}, {Fouesneau},
  {Demleitner}, \& {Andrae}}]{bailer-jones_2021edr3}
{Bailer-Jones}, C.~A.~L., {Rybizki}, J., {Fouesneau}, M., {Demleitner}, M., \&
  {Andrae}, R. 2021, \aj, 161, 147

\bibitem[{{Baraffe} {et~al.}(2015){Baraffe}, {Homeier}, {Allard}, \&
  {Chabrier}}]{Baraffe2015}
{Baraffe}, I., {Homeier}, D., {Allard}, F., \& {Chabrier}, G. 2015, \aap, 577,
  A42

\bibitem[{{Bayo} {et~al.}(2008){Bayo}, {Rodrigo}, {Barrado Y Navascu{\'e}s},
  {Solano}, {Guti{\'e}rrez}, {Morales-Calder{\'o}n}, \& {Allard}}]{bayo2008}
{Bayo}, A., {Rodrigo}, C., {Barrado Y Navascu{\'e}s}, D., {et~al.} 2008, \aap,
  492, 277

\bibitem[{{B{\'e}dard} {et~al.}(2020){B{\'e}dard}, {Bergeron}, {Brassard}, \&
  {Fontaine}}]{Bedard2020}
{B{\'e}dard}, A., {Bergeron}, P., {Brassard}, P., \& {Fontaine}, G. 2020, \apj,
  901, 93

\bibitem[{{Belokurov} {et~al.}(2020){Belokurov}, {Penoyre}, {Oh}, {Iorio},
  {Hodgkin}, {Evans}, {Everall}, {Koposov}, {Tout}, {Izzard}, {Clarke}, \&
  {Brown}}]{Belokurov2020}
{Belokurov}, V., {Penoyre}, Z., {Oh}, S., {et~al.} 2020, \mnras, 496, 1922

\bibitem[{{Bobrick} {et~al.}(2017){Bobrick}, {Davies}, \&
  {Church}}]{Bobrick_2017}
{Bobrick}, A., {Davies}, M.~B., \& {Church}, R.~P. 2017, \mnras, 467, 3556

\bibitem[{{Boller} {et~al.}(2016){Boller}, {Freyberg}, {Tr{\"u}mper}, {Haberl},
  {Voges}, \& {Nandra}}]{Boller_2016_2RXS}
{Boller}, T., {Freyberg}, M.~J., {Tr{\"u}mper}, J., {et~al.} 2016, \aap, 588,
  A103

\bibitem[{{Boro Saikia} {et~al.}(2018){Boro Saikia}, {Marvin}, {Jeffers},
  {Reiners}, {Cameron}, {Marsden}, {Petit}, {Warnecke}, \&
  {Yadav}}]{Boro_Saikia_2018_active_MS}
{Boro Saikia}, S., {Marvin}, C.~J., {Jeffers}, S.~V., {et~al.} 2018, \aap, 616,
  A108

\bibitem[{{Bovy} {et~al.}(2016){Bovy}, {Rix}, {Green}, {Schlafly}, \&
  {Finkbeiner}}]{Bovy_2016_mwdust}
{Bovy}, J., {Rix}, H.-W., {Green}, G.~M., {Schlafly}, E.~F., \& {Finkbeiner},
  D.~P. 2016, \apj, 818, 130

\bibitem[{{Brasseur} {et~al.}(2019){Brasseur}, {Osten}, \&
  {Fleming}}]{Brasseur_2019_flares}
{Brasseur}, C.~E., {Osten}, R.~A., \& {Fleming}, S.~W. 2019, \apj, 883, 88

\bibitem[{{Camacho} {et~al.}(2014){Camacho}, {Torres}, {Garc{\'\i}a-Berro},
  {Zorotovic}, {Schreiber}, {Rebassa-Mansergas}, {Nebot G{\'o}mez-Mor{\'a}n},
  \& {G{\"a}nsicke}}]{Camacho2014}
{Camacho}, J., {Torres}, S., {Garc{\'\i}a-Berro}, E., {et~al.} 2014, \aap, 566,
  A86

\bibitem[{{Chen} {et~al.}(2012){Chen}, {Hou}, {Yu}, {Liu}, {Deng}, {Newberg},
  {Carlin}, {Yang}, {Zhang}, {Shen}, {Zhang}, {Chen}, {Chen}, {Christlieb},
  {Han}, {Lee}, {Liu}, {Pan}, {Shi}, {Wang}, \& {Zhu}}]{Chen2012_LEGUE_LAMOST}
{Chen}, L., {Hou}, J.-L., {Yu}, J.-C., {et~al.} 2012, Research in Astronomy and
  Astrophysics, 12, 805

\bibitem[{{Choi} {et~al.}(2016){Choi}, {Dotter}, {Conroy}, {Cantiello},
  {Paxton}, \& {Johnson}}]{mist1}
{Choi}, J., {Dotter}, A., {Conroy}, C., {et~al.} 2016, \apj, 823, 102

\bibitem[{{Cote} {et~al.}(2019){Cote}, {Abraham}, {Balogh}, {Capak},
  {Carlberg}, {Cowan}, {Djazovski}, {Drissen}, {Drout}, {Dupuis}, {Evans},
  {Fantin}, {Ferrarese}, {Fraser}, {Gallagher}, {Girard}, {Gleisinger},
  {Grandmont}, {Hall}, {Hellmich}, {Hardy}, {Harrison}, {Hlozek}, {Haggard},
  {Henault-Brunet}, {Hutchings}, {Khatu}, {Kavelaars}, {Laurin}, {Lavigne},
  {Lisman}, {Marois}, {McCabe}, {Metchev}, {Moutard}, {Netterfield}, {Nikzad},
  {Ouellette}, {Pass}, {Parker}, {Pazder}, {Percival}, {Rhodes}, {Robert},
  {Rowe}, {Sanchez-Janssen}, {Sivakoff}, {Shapiro}, {Sawicki}, {Scott}, {Van
  Waerbeke}, \& {Venn}}]{Cote_2019_CASTOR}
{Cote}, P., {Abraham}, B., {Balogh}, M., {et~al.} 2019, in Canadian Long Range
  Plan for Astronomy and Astrophysics White Papers, Vol. 2020, 18

\bibitem[{{Cui} {et~al.}(2012){Cui}, {Zhao}, {Chu}, {Li}, {Li}, {Zhang}, {Su},
  {Yao}, {Wang}, {Xing}, {Li}, {Zhu}, {Wang}, {Gu}, {Luo}, {Xu}, {Zhang},
  {Liu}, {Zhang}, {Yang}, {Cao}, {Chen}, {Chen}, {Chen}, {Chen}, {Chu}, {Feng},
  {Gong}, {Hou}, {Hu}, {Hu}, {Hu}, {Jia}, {Jiang}, {Jiang}, {Jiang}, {Jin},
  {Li}, {Li}, {Li}, {Liu}, {Liu}, {Lu}, {Mao}, {Men}, {Qi}, {Qi}, {Shi},
  {Tang}, {Tao}, {Wang}, {Wang}, {Wang}, {Wang}, {Wang}, {Wang}, {Wang},
  {Wang}, {Wang}, {Wang}, {Wang}, {Wang}, {Xu}, {Xu}, {Yang}, {Yu}, {Yuan},
  {Yuan}, {Zhai}, {Zhang}, {Zhang}, {Zhang}, {Zhao}, {Zhou}, {Zhou}, {Zhu}, \&
  {Zou}}]{Cui2012_LAMOST}
{Cui}, X.-Q., {Zhao}, Y.-H., {Chu}, Y.-Q., {et~al.} 2012, Research in Astronomy
  and Astrophysics, 12, 1197

\bibitem[{{Dotter}(2016)}]{mist0}
{Dotter}, A. 2016, \apjs, 222, 8

\bibitem[{{Drimmel} {et~al.}(2003){Drimmel}, {Cabrera-Lavers}, \&
  {L{\'o}pez-Corredoira}}]{Drimmel_2003_mwdust}
{Drimmel}, R., {Cabrera-Lavers}, A., \& {L{\'o}pez-Corredoira}, M. 2003, \aap,
  409, 205

\bibitem[{{Eisenstein} {et~al.}(2011){Eisenstein}, {Weinberg}, {Agol},
  {Aihara}, {Allende Prieto}, {Anderson}, {Arns}, {Aubourg}, {Bailey},
  {Balbinot}, {Barkhouser}, {Beers}, {Berlind}, {Bickerton}, {Bizyaev},
  {Blanton}, {Bochanski}, {Bolton}, {Bosman}, {Bovy}, {Brandt}, {Breslauer},
  {Brewington}, {Brinkmann}, {Brown}, {Brownstein}, {Burger}, {Busca},
  {Campbell}, {Cargile}, {Carithers}, {Carlberg}, {Carr}, {Chang}, {Chen},
  {Chiappini}, {Comparat}, {Connolly}, {Cortes}, {Croft}, {Cunha}, {da Costa},
  {Davenport}, {Dawson}, {De Lee}, {Porto de Mello}, {de Simoni}, {Dean},
  {Dhital}, {Ealet}, {Ebelke}, {Edmondson}, {Eiting}, {Escoffier}, {Esposito},
  {Evans}, {Fan}, {Femen{\'\i}a Castell{\'a}}, {Dutra Ferreira}, {Fitzgerald},
  {Fleming}, {Font-Ribera}, {Ford}, {Frinchaboy}, {Garc{\'\i}a P{\'e}rez},
  {Gaudi}, {Ge}, {Ghezzi}, {Gillespie}, {Gilmore}, {Girardi}, {Gott}, {Gould},
  {Grebel}, {Gunn}, {Hamilton}, {Harding}, {Harris}, {Hawley}, {Hearty},
  {Hennawi}, {Gonz{\'a}lez Hern{\'a}ndez}, {Ho}, {Hogg}, {Holtzman},
  {Honscheid}, {Inada}, {Ivans}, {Jiang}, {Jiang}, {Johnson}, {Jordan},
  {Jordan}, {Kauffmann}, {Kazin}, {Kirkby}, {Klaene}, {Knapp}, {Kneib},
  {Kochanek}, {Koesterke}, {Kollmeier}, {Kron}, {Lampeitl}, {Lang}, {Lawler},
  {Le Goff}, {Lee}, {Lee}, {Leisenring}, {Lin}, {Liu}, {Long}, {Loomis},
  {Lucatello}, {Lundgren}, {Lupton}, {Ma}, {Ma}, {MacDonald}, {Mack},
  {Mahadevan}, {Maia}, {Majewski}, {Makler}, {Malanushenko}, {Malanushenko},
  {Mandelbaum}, {Maraston}, {Margala}, {Maseman}, {Masters}, {McBride},
  {McDonald}, {McGreer}, {McMahon}, {Mena Requejo}, {M{\'e}nard},
  {Miralda-Escud{\'e}}, {Morrison}, {Mullally}, {Muna}, {Murayama}, {Myers},
  {Naugle}, {Neto}, {Nguyen}, {Nichol}, {Nidever}, {O'Connell}, {Ogando},
  {Olmstead}, {Oravetz}, {Padmanabhan}, {Paegert}, {Palanque-Delabrouille},
  {Pan}, {Pandey}, {Parejko}, {P{\^a}ris}, {Pellegrini}, {Pepper}, {Percival},
  {Petitjean}, {Pfaffenberger}, {Pforr}, {Phleps}, {Pichon}, {Pieri}, {Prada},
  {Price-Whelan}, {Raddick}, {Ramos}, {Reid}, {Reyle}, {Rich}, {Richards},
  {Rieke}, {Rieke}, {Rix}, {Robin}, {Rocha-Pinto}, {Rockosi}, {Roe},
  {Rollinde}, {Ross}, {Ross}, {Rossetto}, {S{\'a}nchez}, {Santiago}, {Sayres},
  {Schiavon}, {Schlegel}, {Schlesinger}, {Schmidt}, {Schneider}, {Sellgren},
  {Shelden}, {Sheldon}, {Shetrone}, {Shu}, {Silverman}, {Simmerer}, {Simmons},
  {Sivarani}, {Skrutskie}, {Slosar}, {Smee}, {Smith}, {Snedden}, {Stassun},
  {Steele}, {Steinmetz}, {Stockett}, {Stollberg}, {Strauss}, {Szalay},
  {Tanaka}, {Thakar}, {Thomas}, {Tinker}, {Tofflemire}, {Tojeiro}, {Tremonti},
  {Vargas Maga{\~n}a}, {Verde}, {Vogt}, {Wake}, {Wan}, {Wang}, {Weaver},
  {White}, {White}, {Wilson}, {Wisniewski}, {Wood-Vasey}, {Yanny}, {Yasuda},
  {Y{\`e}che}, {York}, {Young}, {Zasowski}, {Zehavi}, \&
  {Zhao}}]{Eisenstein2011_sdssIII}
{Eisenstein}, D.~J., {Weinberg}, D.~H., {Agol}, E., {et~al.} 2011, \aj, 142, 72

\bibitem[{{Evans} {et~al.}(2018){Evans}, {Riello}, {De Angeli}, {Carrasco},
  {Montegriffo}, {Fabricius}, {Jordi}, {Palaversa}, {Diener}, {Busso},
  {Cacciari}, {van Leeuwen}, {Burgess}, {Davidson}, {Harrison}, {Hodgkin},
  {Pancino}, {Richards}, {Altavilla}, {Balaguer-N{\'u}{\~n}ez}, {Barstow},
  {Bellazzini}, {Brown}, {Castellani}, {Cocozza}, {De Luise}, {Delgado},
  {Ducourant}, {Galleti}, {Gilmore}, {Giuffrida}, {Holl}, {Kewley}, {Koposov},
  {Marinoni}, {Marrese}, {Osborne}, {Piersimoni}, {Portell}, {Pulone},
  {Ragaini}, {Sanna}, {Terrett}, {Walton}, {Wevers}, \&
  {Wyrzykowski}}]{Evans2018_gaia_DR2_photometry}
{Evans}, D.~W., {Riello}, M., {De Angeli}, F., {et~al.} 2018, \aap, 616, A4

\bibitem[{{Evans} {et~al.}(2010){Evans}, {Primini}, {Glotfelty}, {Anderson},
  {Bonaventura}, {Chen}, {Davis}, {Doe}, {Evans}, {Fabbiano}, {Galle}, {Gibbs},
  {Grier}, {Hain}, {Hall}, {Harbo}, {He}, {Houck}, {Karovska}, {Kashyap},
  {Lauer}, {McCollough}, {McDowell}, {Miller}, {Mitschang}, {Morgan},
  {Mossman}, {Nichols}, {Nowak}, {Plummer}, {Refsdal}, {Rots}, {Siemiginowska},
  {Sundheim}, {Tibbetts}, {Van Stone}, {Winkelman}, \&
  {Zografou}}]{Evans_2010_CHANDRA}
{Evans}, I.~N., {Primini}, F.~A., {Glotfelty}, K.~J., {et~al.} 2010, \apjs,
  189, 37

\bibitem[{{Farihi} {et~al.}(2010){Farihi}, {Hoard}, \&
  {Wachter}}]{Farihi2010_WD-RD}
{Farihi}, J., {Hoard}, D.~W., \& {Wachter}, S. 2010, \apjs, 190, 275

\bibitem[{{Ferrario}(2012)}]{Ferrario2012}
{Ferrario}, L. 2012, \mnras, 426, 2500

\bibitem[{{Gaia Collaboration} {et~al.}(2023){Gaia Collaboration}, {Arenou},
  {Babusiaux}, {Barstow}, {Faigler}, {Jorissen}, {Kervella}, {Mazeh},
  {Mowlavi}, {Panuzzo}, {Sahlmann}, {Shahaf}, {Sozzetti}, {Bauchet},
  {Damerdji}, {Gavras}, {Giacobbe}, {Gosset}, {Halbwachs}, {Holl}, {Lattanzi},
  {Leclerc}, {Morel}, {Pourbaix}, {Re Fiorentin}, {Sadowski}, {S{\'e}gransan},
  {Siopis}, {Teyssier}, {Zwitter}, {Planquart}, {Brown}, {Vallenari}, {Prusti},
  {de Bruijne}, {Biermann}, {Creevey}, {Ducourant}, {Evans}, {Eyer}, {Guerra},
  {Hutton}, {Jordi}, {Klioner}, {Lammers}, {Lindegren}, {Luri}, {Mignard},
  {Panem}, {Randich}, {Sartoretti}, {Soubiran}, {Tanga}, {Walton},
  {Bailer-Jones}, {Bastian}, {Drimmel}, {Jansen}, {Katz}, {van Leeuwen},
  {Bakker}, {Cacciari}, {Casta{\~n}eda}, {De Angeli}, {Fabricius}, {Fouesneau},
  {Fr{\'e}mat}, {Galluccio}, {Guerrier}, {Heiter}, {Masana}, {Messineo},
  {Nicolas}, {Nienartowicz}, {Pailler}, {Riclet}, {Roux}, {Seabroke}, {Sordo},
  {Th{\'e}venin}, {Gracia-Abril}, {Portell}, {Altmann}, {Andrae}, {Audard},
  {Bellas-Velidis}, {Benson}, {Berthier}, {Blomme}, {Burgess}, {Busonero},
  {Busso}, {C{\'a}novas}, {Carry}, {Cellino}, {Cheek}, {Clementini},
  {Davidson}, {de Teodoro}, {Nu{\~n}ez Campos}, {Delchambre}, {Dell'Oro},
  {Esquej}, {Fern{\'a}ndez-Hern{\'a}ndez}, {Fraile}, {Garabato},
  {Garc{\'\i}a-Lario}, {Haigron}, {Hambly}, {Harrison}, {Hern{\'a}ndez},
  {Hestroffer}, {Hodgkin}, {Jan{\ss}en}, {Jevardat de Fombelle}, {Jordan},
  {Krone-Martins}, {Lanzafame}, {L{\"o}ffler}, {Marchal}, {Marrese},
  {Moitinho}, {Muinonen}, {Osborne}, {Pancino}, {Pauwels}, {Recio-Blanco},
  {Reyl{\'e}}, {Riello}, {Rimoldini}, {Roegiers}, {Rybizki}, {Sarro}, {Smith},
  {Utrilla}, {van Leeuwen}, {Abbas}, {{\'A}brah{\'a}m}, {Abreu Aramburu},
  {Aerts}, {Aguado}, {Ajaj}, {Aldea-Montero}, {Altavilla}, {{\'A}lvarez},
  {Alves}, {Anders}, {Anderson}, {Anglada Varela}, {Antoja}, {Baines}, {Baker},
  {Balaguer-N{\'u}{\~n}ez}, {Balbinot}, {Balog}, {Barache}, {Barbato},
  {Barros}, {Bartolom{\'e}}, {Bassilana}, {Becciani}, {Bellazzini},
  {Berihuete}, {Bernet}, {Bertone}, {Bianchi}, {Binnenfeld}, {Blanco-Cuaresma},
  {Blazere}, {Boch}, {Bombrun}, {Bossini}, {Bouquillon}, {Bragaglia},
  {Bramante}, {Breedt}, {Bressan}, {Brouillet}, {Brugaletta}, {Bucciarelli},
  {Burlacu}, {Butkevich}, {Buzzi}, {Caffau}, {Cancelliere}, {Cantat-Gaudin},
  {Carballo}, {Carlucci}, {Carnerero}, {Carrasco}, {Casamiquela}, {Castellani},
  {Castro-Ginard}, {Chaoul}, {Charlot}, {Chemin}, {Chiaramida}, {Chiavassa},
  {Chornay}, {Comoretto}, {Contursi}, {Cooper}, {Cornez}, {Cowell}, {Crifo},
  {Cropper}, {Crosta}, {Crowley}, {Dafonte}, {Dapergolas}, {David}, {de
  Laverny}, {De Luise}, {De March}, {De Ridder}, {de Souza}, {de Torres}, {del
  Peloso}, {del Pozo}, {Delbo}, {Delgado}, {Delisle}, {Demouchy},
  {Dharmawardena}, {Diakite}, {Diener}, {Distefano}, {Dolding}, {Enke},
  {Fabre}, {Fabrizio}, {Fedorets}, {Fernique}, {Figueras}, {Fournier},
  {Fouron}, {Fragkoudi}, {Gai}, {Garcia-Gutierrez}, {Garcia-Reinaldos},
  {Garc{\'\i}a-Torres}, {Garofalo}, {Gavel}, {Gerlach}, {Geyer}, {Gilmore},
  {Girona}, {Giuffrida}, {Gomel}, {Gomez}, {Gonz{\'a}lez-N{\'u}{\~n}ez},
  {Gonz{\'a}lez-Santamar{\'\i}a}, {Gonz{\'a}lez-Vidal}, {Granvik}, {Guillout},
  {Guiraud}, {Guti{\'e}rrez-S{\'a}nchez}, {Guy}, {Hatzidimitriou}, {Hauser},
  {Haywood}, {Helmer}, {Helmi}, {Sarmiento}, {Hidalgo}, {Hilger},
  {H{\l}adczuk}, {Hobbs}, {Holland}, {Huckle}, {Jardine}, {Jasniewicz},
  {Jean-Antoine Piccolo}, {Jim{\'e}nez-Arranz}, {Juaristi Campillo}, {Julbe},
  {Karbevska}, {Khanna}, {Kordopatis}, {Korn}, {K{\'o}sp{\'a}l},
  {Kostrzewa-Rutkowska}, {Kruszy{\'n}ska}, {Kun}, {Laizeau}, {Lambert},
  {Lanza}, {Lasne}, {Le Campion}, {Lebreton}, {Lebzelter}, {Leccia},
  {Lecoeur-Taibi}, {Liao}, {Licata}, {Lindstr{\o}m}, {Lister}, {Livanou},
  {Lobel}, {Lorca}, {Loup}, {Madrero Pardo}, {Magdaleno Romeo}, {Managau},
  {Mann}, {Manteiga}, {Marchant}, {Marconi}, {Marcos}, {Marcos Santos},
  {Mar{\'\i}n Pina}, {Marinoni}, {Marocco}, {Marshall}, {Martin Polo},
  {Mart{\'\i}n-Fleitas}, {Marton}, {Mary}, {Masip}, {Massari},
  {Mastrobuono-Battisti}, {McMillan}, {Messina}, {Michalik}, {Millar}, {Mints},
  {Molina}, {Molinaro}, {Moln{\'a}r}, {Monari}, {Mongui{\'o}}, {Montegriffo},
  {Montero}, {Mor}, {Mora}, {Morbidelli}, {Morris}, {Muraveva}, {Murphy},
  {Musella}, {Nagy}, {Noval}, {Oca{\~n}a}, {Ogden}, {Ordenovic}, {Osinde},
  {Pagani}, {Pagano}, {Palaversa}, {Palicio}, {Pallas-Quintela}, {Panahi},
  {Payne-Wardenaar}, {Pe{\~n}alosa Esteller}, {Penttil{\"a}}, {Pichon},
  {Piersimoni}, {Pineau}, {Plachy}, {Plum}, {Poggio}, {Pr{\v{s}}a}, {Pulone},
  {Racero}, {Ragaini}, {Rainer}, {Raiteri}, {Ramos}, {Ramos-Lerate}, {Regibo},
  {Richards}, {Rios Diaz}, {Ripepi}, {Riva}, {Rix}, {Rixon}, {Robichon},
  {Robin}, {Robin}, {Roelens}, {Rogues}, {Rohrbasser}, {Romero-G{\'o}mez},
  {Rowell}, {Royer}, {Ruz Mieres}, {Rybicki}, {S{\'a}ez N{\'u}{\~n}ez},
  {Sagrist{\`a} Sell{\'e}s}, {Salguero}, {Samaras}, {Sanchez Gimenez}, {Sanna},
  {Santove{\~n}a}, {Sarasso}, {Schultheis}, {Sciacca}, {Segol}, {Segovia},
  {Semeux}, {Siddiqui}, {Siebert}, {Siltala}, {Silvelo}, {Slezak}, {Slezak},
  {Smart}, {Snaith}, {Solano}, {Solitro}, {Souami}, {Souchay}, {Spagna},
  {Spina}, {Spoto}, {Steele}, {Steidelm{\"u}ller}, {Stephenson}, {S{\"u}veges},
  {Surdej}, {Szabados}, {Szegedi-Elek}, {Taris}, {Taylor}, {Teixeira},
  {Tolomei}, {Tonello}, {Torra}, {Torra}, {Torralba Elipe}, {Trabucchi},
  {Tsounis}, {Turon}, {Ulla}, {Unger}, {Vaillant}, {van Dillen}, {van Reeven},
  {Vanel}, {Vecchiato}, {Viala}, {Vicente}, {Voutsinas}, {Weiler}, {Wevers},
  {Wyrzykowski}, {Yoldas}, {Yvard}, {Zhao}, {Zorec}, \&
  {Zucker}}]{Gaia_DR3_NSS}
{Gaia Collaboration}, {Arenou}, F., {Babusiaux}, C., {et~al.} 2023, \aap, 674,
  A34

\bibitem[{{Ganguly} {et~al.}(2023){Ganguly}, {Nayak}, \&
  {Chatterjee}}]{Ganguly_2023}
{Ganguly}, A., {Nayak}, P.~K., \& {Chatterjee}, S. 2023, \apj, 954, 4

\bibitem[{{Garc{\'\i}a-Berro} {et~al.}(1997){Garc{\'\i}a-Berro}, {Ritossa}, \&
  {Iben}}]{Garcia1997}
{Garc{\'\i}a-Berro}, E., {Ritossa}, C., \& {Iben}, Icko, J. 1997, \apj, 485,
  765

\bibitem[{{Gentile Fusillo} {et~al.}(2021){Gentile Fusillo}, {Tremblay},
  {Cukanovaite}, {Vorontseva}, {Lallement}, {Hollands}, {G{\"a}nsicke},
  {Burdge}, {McCleery}, \& {Jordan}}]{Gentile_2021_Gaia_WD}
{Gentile Fusillo}, N.~P., {Tremblay}, P.~E., {Cukanovaite}, E., {et~al.} 2021,
  \mnras, 508, 3877

\bibitem[{{Gomes da Silva} {et~al.}(2021){Gomes da Silva}, {Santos},
  {Adibekyan}, {Sousa}, {Campante}, {Figueira}, {Bossini}, {Delgado-Mena},
  {Monteiro}, {de Laverny}, {Recio-Blanco}, \& {Lovis}}]{Gomes_2021_active_MS}
{Gomes da Silva}, J., {Santos}, N.~C., {Adibekyan}, V., {et~al.} 2021, \aap,
  646, A77

\bibitem[{{Green} {et~al.}(2019){Green}, {Schlafly}, {Zucker}, {Speagle}, \&
  {Finkbeiner}}]{Green_2019_mwdust}
{Green}, G.~M., {Schlafly}, E., {Zucker}, C., {Speagle}, J.~S., \&
  {Finkbeiner}, D. 2019, \apj, 887, 93

\bibitem[{{Henden} {et~al.}(2015){Henden}, {Levine}, {Terrell}, \&
  {Welch}}]{apass}
{Henden}, A.~A., {Levine}, S., {Terrell}, D., \& {Welch}, D.~L. 2015, in
  American Astronomical Society Meeting Abstracts, Vol. 225, American
  Astronomical Society Meeting Abstracts \#225, 336.16

\bibitem[{{Iben}(1991)}]{Iben_1991_binary_pop}
{Iben}, Icko, J. 1991, \apjs, 76, 55

\bibitem[{{Istrate} {et~al.}(2014{\natexlab{a}}){Istrate}, {Tauris}, \&
  {Langer}}]{istrate_2014_b}
{Istrate}, A.~G., {Tauris}, T.~M., \& {Langer}, N. 2014{\natexlab{a}}, \aap,
  571, A45

\bibitem[{{Istrate} {et~al.}(2014{\natexlab{b}}){Istrate}, {Tauris}, {Langer},
  \& {Antoniadis}}]{istrate_2014_a}
{Istrate}, A.~G., {Tauris}, T.~M., {Langer}, N., \& {Antoniadis}, J.
  2014{\natexlab{b}}, \aap, 571, L3

\bibitem[{{Jackim} {et~al.}(2024){Jackim}, {Heyl}, \&
  {Richer}}]{Jackim_2024_WDMS_galex}
{Jackim}, R., {Heyl}, J., \& {Richer}, H. 2024, arXiv e-prints,
  arXiv:2404.07388

\bibitem[{{Jadhav} {et~al.}(2019){Jadhav}, {Sindhu}, \&
  {Subramaniam}}]{Jadhav_2019}
{Jadhav}, V.~V., {Sindhu}, N., \& {Subramaniam}, A. 2019, \apj, 886, 13

\bibitem[{{Jadhav} {et~al.}(2023){Jadhav}, {Subramaniam}, \&
  {Sagar}}]{Jadhav_2023}
{Jadhav}, V.~V., {Subramaniam}, A., \& {Sagar}, R. 2023, \aap, 676, A47

\bibitem[{{Khurana} {et~al.}(2023){Khurana}, {Chawla}, \&
  {Chatterjee}}]{Khurana_2023}
{Khurana}, A., {Chawla}, C., \& {Chatterjee}, S. 2023, \apj, 949, 102

\bibitem[{{Koester}(2010)}]{Koester2010}
{Koester}, D. 2010, \memsai, 81, 921

\bibitem[{{Kordopatis} {et~al.}(2013){Kordopatis}, {Gilmore}, {Steinmetz},
  {Boeche}, {Seabroke}, {Siebert}, {Zwitter}, {Binney}, {de Laverny},
  {Recio-Blanco}, {Williams}, {Piffl}, {Enke}, {Roeser}, {Bijaoui}, {Wyse},
  {Freeman}, {Munari}, {Carrillo}, {Anguiano}, {Burton}, {Campbell}, {Cass},
  {Fiegert}, {Hartley}, {Parker}, {Reid}, {Ritter}, {Russell}, {Stupar},
  {Watson}, {Bienaym{\'e}}, {Bland-Hawthorn}, {Gerhard}, {Gibson}, {Grebel},
  {Helmi}, {Navarro}, {Conrad}, {Famaey}, {Faure}, {Just}, {Kos},
  {Matijevi{\v{c}}}, {McMillan}, {Minchev}, {Scholz}, {Sharma}, {Siviero}, {de
  Boer}, \& {{\v{Z}}erjal}}]{Kordopatis2013_RAVE_DR4}
{Kordopatis}, G., {Gilmore}, G., {Steinmetz}, M., {et~al.} 2013, \aj, 146, 134

\bibitem[{{Kulkarni} {et~al.}(2021){Kulkarni}, {Harrison}, {Grefenstette},
  {Earnshaw}, {Andreoni}, {Berg}, {Bloom}, {Cenko}, {Chornock}, {Christiansen},
  {Coughlin}, {Wuollet Criswell}, {Darvish}, {Das}, {De}, {Dessart}, {Dixon},
  {Dorsman}, {El-Badry}, {Evans}, {Ford}, {Fremling}, {Gansicke}, {Gezari},
  {Goetberg}, {Green}, {Graham}, {Heida}, {Ho}, {Jaodand}, {Johns-Krull},
  {Kasliwal}, {Lazzarini}, {Lu}, {Margutti}, {Martin}, {Masters}, {McKernan},
  {Naze}, {Nissanke}, {Parazin}, {Perley}, {Phinney}, {Piro}, {Raaijmakers},
  {Rauw}, {Rodriguez}, {Sana}, {Senchyna}, {Singer}, {Spake}, {Stassun},
  {Stern}, {Teplitz}, {Weisz}, \& {Yao}}]{Kulkarni_2021_UVEX}
{Kulkarni}, S.~R., {Harrison}, F.~A., {Grefenstette}, B.~W., {et~al.} 2021,
  arXiv e-prints, arXiv:2111.15608

\bibitem[{{Kunder} {et~al.}(2017){Kunder}, {Kordopatis}, {Steinmetz},
  {Zwitter}, {McMillan}, {Casagrande}, {Enke}, {Wojno}, {Valentini},
  {Chiappini}, {Matijevi{\v{c}}}, {Siviero}, {de Laverny}, {Recio-Blanco},
  {Bijaoui}, {Wyse}, {Binney}, {Grebel}, {Helmi}, {Jofre}, {Antoja}, {Gilmore},
  {Siebert}, {Famaey}, {Bienaym{\'e}}, {Gibson}, {Freeman}, {Navarro},
  {Munari}, {Seabroke}, {Anguiano}, {{\v{Z}}erjal}, {Minchev}, {Reid},
  {Bland-Hawthorn}, {Kos}, {Sharma}, {Watson}, {Parker}, {Scholz}, {Burton},
  {Cass}, {Hartley}, {Fiegert}, {Stupar}, {Ritter}, {Hawkins}, {Gerhard},
  {Chaplin}, {Davies}, {Elsworth}, {Lund}, {Miglio}, \&
  {Mosser}}]{Kunder2017_RAVE_DR5}
{Kunder}, A., {Kordopatis}, G., {Steinmetz}, M., {et~al.} 2017, \aj, 153, 75

\bibitem[{{Li} {et~al.}(2025){Li}, {Ting}, {Rix}, {Green}, {Hogg}, {Ren},
  {M{\"u}ller-Horn}, \& {Seeburger}}]{Li_2025_30k_wdms}
{Li}, J., {Ting}, Y.-S., {Rix}, H.-W., {et~al.} 2025, \apjs, 279, 47

\bibitem[{{Li} {et~al.}(2024){Li}, {Wang}, {Han}, {Yang}, {Zheng}, {Huang}, \&
  {Liu}}]{Li_2024_flares}
{Li}, X., {Wang}, S., {Han}, H., {et~al.} 2024, \apj, 966, 69

\bibitem[{{Magnier} {et~al.}(2020){Magnier}, {Schlafly}, {Finkbeiner}, {Tonry},
  {Goldman}, {R{\"o}ser}, {Schilbach}, {Casertano}, {Chambers}, {Flewelling},
  {Huber}, {Price}, {Sweeney}, {Waters}, {Denneau}, {Draper}, {Hodapp},
  {Jedicke}, {Kaiser}, {Kudritzki}, {Metcalfe}, {Stubbs}, \&
  {Wainscoat}}]{magnier_panstarrs_dr2}
{Magnier}, E.~A., {Schlafly}, E.~F., {Finkbeiner}, D.~P., {et~al.} 2020, \apjs,
  251, 6

\bibitem[{{Marshall} {et~al.}(2006){Marshall}, {Robin}, {Reyl{\'e}},
  {Schultheis}, \& {Picaud}}]{Marshall_2006_mwdust}
{Marshall}, D.~J., {Robin}, A.~C., {Reyl{\'e}}, C., {Schultheis}, M., \&
  {Picaud}, S. 2006, \aap, 453, 635

\bibitem[{{Martin} {et~al.}(2005){Martin}, {Fanson}, {Schiminovich},
  {Morrissey}, {Friedman}, {Barlow}, {Conrow}, {Grange}, {Jelinsky},
  {Milliard}, {Siegmund}, {Bianchi}, {Byun}, {Donas}, {Forster}, {Heckman},
  {Lee}, {Madore}, {Malina}, {Neff}, {Rich}, {Small}, {Surber}, {Szalay},
  {Welsh}, \& {Wyder}}]{martin2005_GALEX}
{Martin}, D.~C., {Fanson}, J., {Schiminovich}, D., {et~al.} 2005, \apjl, 619,
  L1

\bibitem[{{Mart{\'\i}nez-Arn{\'a}iz} {et~al.}(2010){Mart{\'\i}nez-Arn{\'a}iz},
  {Maldonado}, {Montes}, {Eiroa}, \& {Montesinos}}]{Martinez_2010_active_MS}
{Mart{\'\i}nez-Arn{\'a}iz}, R., {Maldonado}, J., {Montes}, D., {Eiroa}, C., \&
  {Montesinos}, B. 2010, \aap, 520, A79

\bibitem[{{Medina} {et~al.}(2020){Medina}, {Winters}, {Irwin}, \&
  {Charbonneau}}]{Medina_2020_flares_15pc}
{Medina}, A.~A., {Winters}, J.~G., {Irwin}, J.~M., \& {Charbonneau}, D. 2020,
  \apj, 905, 107

\bibitem[{{Michalik} {et~al.}(2015){Michalik}, {Lindegren}, \&
  {Hobbs}}]{Michalik2015_TGAS}
{Michalik}, D., {Lindegren}, L., \& {Hobbs}, D. 2015, \aap, 574, A115

\bibitem[{{Morgan} {et~al.}(2012){Morgan}, {West}, {Garc{\'e}s}, {Catal{\'a}n},
  {Dhital}, {Fuchs}, \& {Silvestri}}]{morgan2012}
{Morgan}, D.~P., {West}, A.~A., {Garc{\'e}s}, A., {et~al.} 2012, \aj, 144, 93

\bibitem[{{Muirhead} {et~al.}(2013){Muirhead}, {Vanderburg}, {Shporer},
  {Becker}, {Swift}, {Lloyd}, {Fuller}, {Zhao}, {Hinkley}, {Pineda}, {Bottom},
  {Howard}, {von Braun}, {Boyajian}, {Law}, {Baranec}, {Riddle}, {Ramaprakash},
  {Tendulkar}, {Bui}, {Burse}, {Chordia}, {Das}, {Dekany}, {Punnadi}, \&
  {Johnson}}]{Muirhead_2013_post_CE_Mdwarf}
{Muirhead}, P.~S., {Vanderburg}, A., {Shporer}, A., {et~al.} 2013, \apj, 767,
  111

\bibitem[{{Nandez} {et~al.}(2015){Nandez}, {Ivanova}, \&
  {Lombardi}}]{nandez_2015}
{Nandez}, J.~L.~A., {Ivanova}, N., \& {Lombardi}, J.~C.~J. 2015, \mnras, 450,
  L39

\bibitem[{{Nayak} {et~al.}(2024){Nayak}, {Ganguly}, \&
  {Chatterjee}}]{nayak_2024_wdms}
{Nayak}, P.~K., {Ganguly}, A., \& {Chatterjee}, S. 2024, \mnras, 527, 6100

\bibitem[{{Newton} {et~al.}(2017){Newton}, {Irwin}, {Charbonneau}, {Berlind},
  {Calkins}, \& {Mink}}]{Newton_2017_rotation_flares}
{Newton}, E.~R., {Irwin}, J., {Charbonneau}, D., {et~al.} 2017, \apj, 834, 85

\bibitem[{{Parsons} {et~al.}(2017){Parsons}, {G{\"a}nsicke}, {Marsh}, {Ashley},
  {Bours}, {Breedt}, {Burleigh}, {Copperwheat}, {Dhillon}, {Green}, {Hardy},
  {Hermes}, {Irawati}, {Kerry}, {Littlefair}, {McAllister}, {Rattanasoon},
  {Rebassa-Mansergas}, {Sahman}, \& {Schreiber}}]{Parsons2017}
{Parsons}, S.~G., {G{\"a}nsicke}, B.~T., {Marsh}, T.~R., {et~al.} 2017, \mnras,
  470, 4473

\bibitem[{{Parsons} {et~al.}(2016){Parsons}, {Rebassa-Mansergas}, {Schreiber},
  {G{\"a}nsicke}, {Zorotovic}, \& {Ren}}]{WDS1_parsons2016}
{Parsons}, S.~G., {Rebassa-Mansergas}, A., {Schreiber}, M.~R., {et~al.} 2016,
  \mnras, 463, 2125

\bibitem[{{Pecaut} \& {Mamajek}(2013)}]{pecaut2013}
{Pecaut}, M.~J. \& {Mamajek}, E.~E. 2013, \apjs, 208, 9

\bibitem[{{P{\'e}rez-Couto} {et~al.}(2025){P{\'e}rez-Couto}, {Manteiga}, \&
  {Villaver}}]{Xabier_2025_ML_WDMS}
{P{\'e}rez-Couto}, X., {Manteiga}, M., \& {Villaver}, E. 2025, \apj, 988, 51

\bibitem[{{Rebassa-Mansergas} {et~al.}(2013{\natexlab{a}}){Rebassa-Mansergas},
  {Agurto-Gangas}, {Schreiber}, {G{\"a}nsicke}, \& {Koester}}]{Rebassa2013b}
{Rebassa-Mansergas}, A., {Agurto-Gangas}, C., {Schreiber}, M.~R.,
  {G{\"a}nsicke}, B.~T., \& {Koester}, D. 2013{\natexlab{a}}, \mnras, 433, 3398

\bibitem[{{Rebassa-Mansergas} {et~al.}(2016{\natexlab{a}}){Rebassa-Mansergas},
  {Anguiano}, {Garc{\'\i}a-Berro}, {Freeman}, {Cojocaru}, {Manser}, {Pala},
  {G{\"a}nsicke}, \& {Liu}}]{Rebassa2016b}
{Rebassa-Mansergas}, A., {Anguiano}, B., {Garc{\'\i}a-Berro}, E., {et~al.}
  2016{\natexlab{a}}, \mnras, 463, 1137

\bibitem[{{Rebassa-Mansergas} {et~al.}(2010){Rebassa-Mansergas},
  {G{\"a}nsicke}, {Schreiber}, {Koester}, \&
  {Rodr{\'\i}guez-Gil}}]{Rebassa_2010_wd_sdssVII}
{Rebassa-Mansergas}, A., {G{\"a}nsicke}, B.~T., {Schreiber}, M.~R., {Koester},
  D., \& {Rodr{\'\i}guez-Gil}, P. 2010, \mnras, 402, 620

\bibitem[{{Rebassa-Mansergas} {et~al.}(2012{\natexlab{a}}){Rebassa-Mansergas},
  {Nebot G{\'o}mez-Mor{\'a}n}, {Schreiber}, {G{\"a}nsicke}, {Schwope},
  {Gallardo}, \& {Koester}}]{Rebassa2012a}
{Rebassa-Mansergas}, A., {Nebot G{\'o}mez-Mor{\'a}n}, A., {Schreiber}, M.~R.,
  {et~al.} 2012{\natexlab{a}}, \mnras, 419, 806

\bibitem[{{Rebassa-Mansergas} {et~al.}(2017){Rebassa-Mansergas}, {Ren},
  {Irawati}, {Garc{\'\i}a-Berro}, {Parsons}, {Schreiber}, {G{\"a}nsicke},
  {Rodr{\'\i}guez-Gil}, {Liu}, {Manser}, {Nevado}, {Jim{\'e}nez-Ibarra},
  {Costero}, {Echevarr{\'\i}a}, {Michel}, {Zorotovic}, {Hollands}, {Han},
  {Luo}, {Villaver}, \& {Kong}}]{WDS2_Rebassa2017}
{Rebassa-Mansergas}, A., {Ren}, J.~J., {Irawati}, P., {et~al.} 2017, \mnras,
  472, 4193

\bibitem[{{Rebassa-Mansergas} {et~al.}(2016{\natexlab{b}}){Rebassa-Mansergas},
  {Ren}, {Parsons}, {G{\"a}nsicke}, {Schreiber}, {Garc{\'\i}a-Berro}, {Liu}, \&
  {Koester}}]{Rebassa2016a}
{Rebassa-Mansergas}, A., {Ren}, J.~J., {Parsons}, S.~G., {et~al.}
  2016{\natexlab{b}}, \mnras, 458, 3808

\bibitem[{{Rebassa-Mansergas} {et~al.}(2013{\natexlab{b}}){Rebassa-Mansergas},
  {Schreiber}, \& {G{\"a}nsicke}}]{Rebassa2013a}
{Rebassa-Mansergas}, A., {Schreiber}, M.~R., \& {G{\"a}nsicke}, B.~T.
  2013{\natexlab{b}}, \mnras, 429, 3570

\bibitem[{{Rebassa-Mansergas} {et~al.}(2025){Rebassa-Mansergas}, {Solano},
  {Brown}, {Parsons}, {Murillo-Ojeda}, {Raddi}, {Camisassa}, {Torres}, \& {van
  Roestel}}]{Rebassa_2025_mag_limited}
{Rebassa-Mansergas}, A., {Solano}, E., {Brown}, A.~J., {et~al.} 2025, \aap,
  699, A153

\bibitem[{{Rebassa-Mansergas} {et~al.}(2021){Rebassa-Mansergas}, {Solano},
  {Jim{\'e}nez-Esteban}, {Torres}, {Rodrigo}, {Ferrer-Burjachs}, {Calcaferro},
  {Althaus}, \& {C{\'o}rsico}}]{Rebassa2021}
{Rebassa-Mansergas}, A., {Solano}, E., {Jim{\'e}nez-Esteban}, F.~M., {et~al.}
  2021, \mnras, 506, 5201

\bibitem[{{Rebassa-Mansergas} {et~al.}(2012{\natexlab{b}}){Rebassa-Mansergas},
  {Zorotovic}, {Schreiber}, {G{\"a}nsicke}, {Southworth}, {Nebot
  G{\'o}mez-Mor{\'a}n}, {Tappert}, {Koester}, {Pyrzas}, {Papadaki},
  {Schmidtobreick}, {Schwope}, \& {Toloza}}]{Rebassa2012b}
{Rebassa-Mansergas}, A., {Zorotovic}, M., {Schreiber}, M.~R., {et~al.}
  2012{\natexlab{b}}, \mnras, 423, 320

\bibitem[{{Rekhi} {et~al.}(2023){Rekhi}, {Ben-Ami}, {Perdelwitz}, \&
  {Shvartzvald}}]{Rekhi_2023_flares}
{Rekhi}, P., {Ben-Ami}, S., {Perdelwitz}, V., \& {Shvartzvald}, Y. 2023, \apj,
  955, 24

\bibitem[{{Ren} {et~al.}(2020){Ren}, {Raddi}, {Rebassa-Mansergas}, {Hernandez},
  {Parsons}, {Irawati}, {Rittipruk}, {Schreiber}, {G{\"a}nsicke}, {Torres},
  {Wang}, {Zhang}, {Zhao}, {Zhou}, {Han}, {Wang}, {Liu}, {Liu}, {Wang},
  {Zheng}, {Wang}, {Zhao}, {Cui}, {Shi}, \& {Tian}}]{WDS5_Ren2020}
{Ren}, J.~J., {Raddi}, R., {Rebassa-Mansergas}, A., {et~al.} 2020, \apj, 905,
  38

\bibitem[{{Ren} {et~al.}(2018){Ren}, {Rebassa-Mansergas}, {Parsons}, {Liu},
  {Luo}, {Kong}, \& {Zhang}}]{Ren2018_wdms_LAMOST_DR5}
{Ren}, J.~J., {Rebassa-Mansergas}, A., {Parsons}, S.~G., {et~al.} 2018, \mnras,
  477, 4641

\bibitem[{{Samus'} {et~al.}(2017){Samus'}, {Kazarovets}, {Durlevich},
  {Kireeva}, \& {Pastukhova}}]{Samus_2017}
{Samus'}, N.~N., {Kazarovets}, E.~V., {Durlevich}, O.~V., {Kireeva}, N.~N., \&
  {Pastukhova}, E.~N. 2017, Astronomy Reports, 61, 80

\bibitem[{{Saxton} {et~al.}(2008){Saxton}, {Read}, {Esquej}, {Freyberg},
  {Altieri}, \& {Bermejo}}]{Saxton_2008_XMM-Newton-slew}
{Saxton}, R.~D., {Read}, A.~M., {Esquej}, P., {et~al.} 2008, \aap, 480, 611

\bibitem[{{Schlegel} {et~al.}(1998){Schlegel}, {Finkbeiner}, \&
  {Davis}}]{Schlegel1998}
{Schlegel}, D.~J., {Finkbeiner}, D.~P., \& {Davis}, M. 1998, \apj, 500, 525

\bibitem[{{Shahaf} {et~al.}(2023){Shahaf}, {Bashi}, {Mazeh}, {Faigler},
  {Arenou}, {El-Badry}, \& {Rix}}]{Shahaf2023}
{Shahaf}, S., {Bashi}, D., {Mazeh}, T., {et~al.} 2023, \mnras, 518, 2991

\bibitem[{{Sidharth} {et~al.}(2024){Sidharth}, {Shridharan}, {Mathew},
  {Devaraj}, {Cysil}, {Stalin}, {Arun}, {Bhattacharyya}, {Kartha}, \&
  {Robin}}]{Sidharth_2024_WD_towards_SMC}
{Sidharth}, A.~V., {Shridharan}, B., {Mathew}, B., {et~al.} 2024, arXiv
  e-prints, arXiv:2408.00852

\bibitem[{{Skinner} {et~al.}(2017){Skinner}, {Morgan}, {West}, {L{\'e}pine}, \&
  {Thorstensen}}]{Skinner2017}
{Skinner}, J.~N., {Morgan}, D.~P., {West}, A.~A., {L{\'e}pine}, S., \&
  {Thorstensen}, J.~R. 2017, \aj, 154, 118

\bibitem[{{Skrutskie} {et~al.}(2006){Skrutskie}, {Cutri}, {Stiening},
  {Weinberg}, {Schneider}, {Carpenter}, {Beichman}, {Capps}, {Chester},
  {Elias}, {Huchra}, {Liebert}, {Lonsdale}, {Monet}, {Price}, {Seitzer},
  {Jarrett}, {Kirkpatrick}, {Gizis}, {Howard}, {Evans}, {Fowler}, {Fullmer},
  {Hurt}, {Light}, {Kopan}, {Marsh}, {McCallon}, {Tam}, {Van Dyk}, \&
  {Wheelock}}]{2mass}
{Skrutskie}, M.~F., {Cutri}, R.~M., {Stiening}, R., {et~al.} 2006, \aj, 131,
  1163

\bibitem[{{Sriram} {et~al.}(2023){Sriram}, {Valsan}, {Subramaniam}, {Unni},
  {Maheswar}, \& {Chand}}]{Sriram_2023_INSIST}
{Sriram}, S., {Valsan}, V., {Subramaniam}, A., {et~al.} 2023, Journal of
  Astrophysics and Astronomy, 44, 55

\bibitem[{{Subramaniam}(2022)}]{purni_2022_INSIST}
{Subramaniam}, A. 2022, Journal of Astrophysics and Astronomy, 43, 80

\bibitem[{{Subramaniam} {et~al.}(2020){Subramaniam}, {Pandey}, {Jadhav}, \&
  {Sahu}}]{Subramaniam_2020}
{Subramaniam}, A., {Pandey}, S., {Jadhav}, V.~V., \& {Sahu}, S. 2020, Journal
  of Astrophysics and Astronomy, 41, 45

\bibitem[{{Tandon} {et~al.}(2017{\natexlab{a}}){Tandon}, {Hutchings}, {Ghosh},
  {Subramaniam}, {Koshy}, {Girish}, {Kamath}, {Kathiravan}, {Kumar},
  {Lancelot}, {Mahesh}, {Mohan}, {Murthy}, {Nagabhushana}, {Pati}, {Postma},
  {Rao}, {Sankarasubramanian}, {Sreekumar}, {Sriram}, {Stalin}, {Sutaria},
  {Sreedhar}, {Barve}, {Mondal}, \& {Sahu}}]{tandon2017b}
{Tandon}, S.~N., {Hutchings}, J.~B., {Ghosh}, S.~K., {et~al.}
  2017{\natexlab{a}}, Journal of Astrophysics and Astronomy, 38, 28

\bibitem[{{Tandon} {et~al.}(2017{\natexlab{b}}){Tandon}, {Subramaniam},
  {Girish}, {Postma}, {Sankarasubramanian}, {Sriram}, {Stalin}, {Mondal},
  {Sahu}, {Joseph}, {Hutchings}, {Ghosh}, {Barve}, {George}, {Kamath},
  {Kathiravan}, {Kumar}, {Lancelot}, {Leahy}, {Mahesh}, {Mohan},
  {Nagabhushana}, {Pati}, {Kameswara Rao}, {Sreedhar}, \&
  {Sreekumar}}]{tandon2017a}
{Tandon}, S.~N., {Subramaniam}, A., {Girish}, V., {et~al.} 2017{\natexlab{b}},
  \aj, 154, 128

\bibitem[{{Toonen} {et~al.}(2017){Toonen}, {Hollands}, {G{\"a}nsicke}, \&
  {Boekholt}}]{Toonen2017}
{Toonen}, S., {Hollands}, M., {G{\"a}nsicke}, B.~T., \& {Boekholt}, T. 2017,
  \aap, 602, A16

\bibitem[{Torres {et~al.}(2022)Torres, Canals, Jiménez-Esteban,
  Rebassa-Mansergas, \& Solano}]{torres_2022}
Torres, S., Canals, P., Jiménez-Esteban, F.~M., Rebassa-Mansergas, A., \&
  Solano, E. 2022, Monthly Notices of the Royal Astronomical Society, 511, 5462

\bibitem[{{Vall{\'e}e}(2003)}]{Vallee_2003}
{Vall{\'e}e}, J.~P. 2003, \nar, 47, 85

\bibitem[{{Wang} \& {Han}(2012)}]{Wang2012}
{Wang}, B. \& {Han}, Z. 2012, \nar, 56, 122

\bibitem[{{Webb} {et~al.}(2020){Webb}, {Coriat}, {Traulsen}, {Ballet}, {Motch},
  {Carrera}, {Koliopanos}, {Authier}, {de la Calle}, {Ceballos}, {Colomo},
  {Chuard}, {Freyberg}, {Garcia}, {Kolehmainen}, {Lamer}, {Lin}, {Maggi},
  {Michel}, {Page}, {Page}, {Perea-Calderon}, {Pineau}, {Rodriguez}, {Rosen},
  {Santos Lleo}, {Saxton}, {Schwope}, {Tom{\'a}s}, {Watson}, \&
  {Zakardjian}}]{Webb_2020_XMM-Newton-epic}
{Webb}, N.~A., {Coriat}, M., {Traulsen}, I., {et~al.} 2020, \aap, 641, A136

\bibitem[{{Wright} {et~al.}(2010){Wright}, {Eisenhardt}, {Mainzer}, {Ressler},
  {Cutri}, {Jarrett}, {Kirkpatrick}, {Padgett}, {McMillan}, {Skrutskie},
  {Stanford}, {Cohen}, {Walker}, {Mather}, {Leisawitz}, {Gautier}, {McLean},
  {Benford}, {Lonsdale}, {Blain}, {Mendez}, {Irace}, {Duval}, {Liu}, {Royer},
  {Heinrichsen}, {Howard}, {Shannon}, {Kendall}, {Walsh}, {Larsen}, {Cardon},
  {Schick}, {Schwalm}, {Abid}, {Fabinsky}, {Naes}, \& {Tsai}}]{allwise}
{Wright}, E.~L., {Eisenhardt}, P. R.~M., {Mainzer}, A.~K., {et~al.} 2010, \aj,
  140, 1868

\bibitem[{{Yang} {et~al.}(2017){Yang}, {Liu}, {Gao}, {Fang}, {Guo}, {Zhang},
  {Hou}, {Wang}, \& {Cao}}]{Yang_2017_flares_kepler_field}
{Yang}, H., {Liu}, J., {Gao}, Q., {et~al.} 2017, \apj, 849, 36

\bibitem[{{York} {et~al.}(2000){York}, {Adelman}, {Anderson}, {Anderson},
  {Annis}, {Bahcall}, {Bakken}, {Barkhouser}, {Bastian}, {Berman}, {Boroski},
  {Bracker}, {Briegel}, {Briggs}, {Brinkmann}, {Brunner}, {Burles}, {Carey},
  {Carr}, {Castander}, {Chen}, {Colestock}, {Connolly}, {Crocker}, {Csabai},
  {Czarapata}, {Davis}, {Doi}, {Dombeck}, {Eisenstein}, {Ellman}, {Elms},
  {Evans}, {Fan}, {Federwitz}, {Fiscelli}, {Friedman}, {Frieman}, {Fukugita},
  {Gillespie}, {Gunn}, {Gurbani}, {de Haas}, {Haldeman}, {Harris}, {Hayes},
  {Heckman}, {Hennessy}, {Hindsley}, {Holm}, {Holmgren}, {Huang}, {Hull},
  {Husby}, {Ichikawa}, {Ichikawa}, {Ivezi{\'c}}, {Kent}, {Kim}, {Kinney},
  {Klaene}, {Kleinman}, {Kleinman}, {Knapp}, {Korienek}, {Kron}, {Kunszt},
  {Lamb}, {Lee}, {Leger}, {Limmongkol}, {Lindenmeyer}, {Long}, {Loomis},
  {Loveday}, {Lucinio}, {Lupton}, {MacKinnon}, {Mannery}, {Mantsch}, {Margon},
  {McGehee}, {McKay}, {Meiksin}, {Merelli}, {Monet}, {Munn}, {Narayanan},
  {Nash}, {Neilsen}, {Neswold}, {Newberg}, {Nichol}, {Nicinski}, {Nonino},
  {Okada}, {Okamura}, {Ostriker}, {Owen}, {Pauls}, {Peoples}, {Peterson},
  {Petravick}, {Pier}, {Pope}, {Pordes}, {Prosapio}, {Rechenmacher}, {Quinn},
  {Richards}, {Richmond}, {Rivetta}, {Rockosi}, {Ruthmansdorfer}, {Sandford},
  {Schlegel}, {Schneider}, {Sekiguchi}, {Sergey}, {Shimasaku}, {Siegmund},
  {Smee}, {Smith}, {Snedden}, {Stone}, {Stoughton}, {Strauss}, {Stubbs},
  {SubbaRao}, {Szalay}, {Szapudi}, {Szokoly}, {Thakar}, {Tremonti}, {Tucker},
  {Uomoto}, {Vanden Berk}, {Vogeley}, {Waddell}, {Wang}, {Watanabe},
  {Weinberg}, {Yanny}, {Yasuda}, \& {SDSS Collaboration}}]{York2000_sdss}
{York}, D.~G., {Adelman}, J., {Anderson}, John~E., J., {et~al.} 2000, \aj, 120,
  1579

\bibitem[{{Zorotovic} {et~al.}(2010){Zorotovic}, {Schreiber}, {G{\"a}nsicke},
  \& {Nebot G{\'o}mez-Mor{\'a}n}}]{Zorotovic2010}
{Zorotovic}, M., {Schreiber}, M.~R., {G{\"a}nsicke}, B.~T., \& {Nebot
  G{\'o}mez-Mor{\'a}n}, A. 2010, \aap, 520, A86

\bibitem[{{Zorotovic} {et~al.}(2014){Zorotovic}, {Schreiber},
  {Garc{\'\i}a-Berro}, {Camacho}, {Torres}, {Rebassa-Mansergas}, \&
  {G{\"a}nsicke}}]{Zorotovic2014}
{Zorotovic}, M., {Schreiber}, M.~R., {Garc{\'\i}a-Berro}, E., {et~al.} 2014,
  \aap, 568, A68

\end{thebibliography}

\appendix

\section{A representative table of online WDMS catalog}

\begin{sidewaystable}
  \caption{The catalog of WDMS binaries and their stellar properties. }
  \tabcolsep1.70pt $ $
  \begin{tabular}{c|c|c|ccc|cccc|c|c|c|c|c|}
    \hline
  & \gaia\ ID & \galex\ ID &  &   MS Properties&  &  & & WD Properties& & $\chired$ & Vgf$_b$  & \% data & $f_{res}$ & $f_{res}$ \\
  \cline{4-10}
  & &  &  $\teff$ & Radius & $\lbol$ &$\teff$& 
 Radius& $\lbol$& Mass & &  & $f_{res}<0.5$ & NUV  & Bluest\\
  &  &  
  & ($10 ^{3}$K) & $(R_{\odot})$ & $(10^{-2} \times L_{\odot})$ & ($10^{3}$K) & 
 $(10^{-3}R_{\odot})$ &$(10^{-3}L_{\odot})$ & $(M_{\odot})$ &  & & Opt-NIR &  & Opt \\
    \hline
1 & 1006621281985546240 & 6374258338484454060 & 3.4$\pm$0.05 & 0.344$\pm$0.001 & 1.47$\pm$0.01 & 13.5$\pm$0.12 & 12.9$\pm$0.02 & 5.19$\pm$0.05 & 0.63$\pm$0.14 & 1.97 & 0.53 & 1.0 & 0.14 & 0.12 \\
2 & 1015888309580557056 & 6373730547133317632 & 3.8$\pm$0.05 & 0.399$\pm$0.001 & 3.52$\pm$0.04 & 11.25$\pm$0.12 & 12.78$\pm$0.04 & 2.64$\pm$0.03 & 0.59$\pm$0.05 & 43.88 & 6.12 & 1.0 & 0.32 & 0.01 \\
3 & 1025122248749370880 & 6373625025153467361 & 3.2$\pm$0.05 & 0.489$\pm$0.001 & 2.22$\pm$0.02 & 7.25$\pm$0.12 & 7.78$\pm$0.02 & 0.15$\pm$0.0 & 1.05$\pm$0.0 & 6.34 & 2.12 & 1.0 & 0.2 & 0.07 \\
4 & 1058663641228841600 & 6375102714022462246 & 3.8$\pm$0.05 & 0.482$\pm$0.001 & 4.56$\pm$0.03 & 10.75$\pm$0.12 & 11.82$\pm$0.01 & 1.64$\pm$0.02 & 0.72$\pm$0.01 & 47.13 & 5.97 & 1.0 & 0.18 & 0.0 \\
5 & 1059280845208783488 & 6375102704358788195 & 3.3$\pm$0.05 & 0.437$\pm$0.001 & 2.09$\pm$0.02 & 13.0$\pm$0.12 & 2.76$\pm$0.0 & 0.22$\pm$0.01 & --- & 5.08 & 10.22 & 0.85 & 0.18 & 0.12 \\
6 & 1089733400290040320 & 6373484278003533137 & 3.1$\pm$0.05 & 0.08$\pm$0.007 & 0.05$\pm$0.01 & 11.0$\pm$0.12 & 3.32$\pm$0.29 & 0.15$\pm$0.03 & --- & 7.5 & 1.67 & 1.0 & 0.12 & 0.2 \\
7 & 1097423796930720896 & 6373449127991182854 & 3.3$\pm$0.05 & 0.088$\pm$0.006 & 0.09$\pm$0.01 & 15.25$\pm$0.12 & 2.56$\pm$0.19 & 0.33$\pm$0.05 & --- & 59.03 & 3.32 & 1.0 & 0.11 & 0.13 \\
8 & 1108145478408876672 & 6373484237199245602 & 2.7$\pm$0.05 & 0.406$\pm$0.001 & 0.87$\pm$0.01 & 5.25$\pm$0.12 & 24.19$\pm$0.06 & 0.46$\pm$0.01 & --- & 4.88 & 4.99 & 0.95 & 0.36 & 0.31 \\
9 & 1156516327809897088 & 6382772924315601554 & 3.1$\pm$0.05 & 0.348$\pm$0.001 & 1.02$\pm$0.01 & 7.75$\pm$0.12 & 12.25$\pm$0.02 & 0.48$\pm$0.0 & 0.72$\pm$0.07 & 4.88 & 0.87 & 1.0 & 0.01 & 0.03 \\
10 & 1169544338008726528 & 6382737725984869051 & 3.1$\pm$0.05 & 0.294$\pm$0.0 & 0.71$\pm$0.0 & 6.0$\pm$0.12 & 8.26$\pm$0.01 & 0.08$\pm$0.0 & 1.02$\pm$0.0 & 58.24 & 3.04 & 0.95 & 0.37 & 0.38 \\
\vdots &  \vdots & \vdots  & \vdots & \vdots & \vdots & \vdots & \vdots & \vdots & \vdots & \vdots & \vdots & \vdots & \vdots \\
    \hline
  \end{tabular}
  \tablefoot{A truncated list of WDMS candidates. The full list is available on \href{https://zenodo.org/records/19270085}{Zenodo} and at the CDS. We provide \gaia\ and \galex\ IDs, MS and WD properties, Vgf$_b$, and fraction of data points in optical-NIR region having $f_{residue} < 0.5$, $f_{residue}$ values in NUV $\&$ in the next bluest optical points to NUV after binary SED fitting.}
\label{tab:source_table}
\end{sidewaystable}

\section{Additional Tables}

\begin{table*}
  \caption{A comparison of estimated stellar parameters of WDMS binaries with LAMOST catalog.}
  \begin{tabular}{|c|c|cc|cc|}
    \hline
  &   &  \multicolumn{2}{c|}{SED parameters}  &  \multicolumn{2}{c|}{ LAMOST parameters }   \\
  \cline{3-6}
 Index &  \gaia\ ID &  WD $\teff$  & MS $\teff$ &  combined $\teff$   & classification \\
  &    
  & ($10 ^{3}$K)  & ($10^{3}$K) & 
 ($10 ^{3}$K)  &   \\
    \hline
1 & 1006621281985546240 & 13.50$\pm$0.12 & 3.4$\pm$0.05 & --- & Double Star  \\
2 & 1015888309580557056 & 11.25$\pm$0.12 & 3.8$\pm$0.05 & 3.66 & MS (M1) \\ 
3 & 1607603170812636928\tablefootmark{a} & 16.75$\pm$0.12 & 3.6$\pm$0.05 & 3.73--3.79 & MS (M1) \\
4 & 1911816151065187456 & 14.75$\pm$0.12 & 3.3$\pm$0.05 & --- & Double Star \\
5 & 245807579721338752 & 8.00$\pm$0.12 & 2.5$\pm$0.05 & --- & WD (DA) \\
6 & 2560009007603950720\tablefootmark{a} & 9.50$\pm$0.12 & 2.7$\pm$0.05 & 3.70--3.80 & DC+M7 \\ 
7 & 3387904051723656320 & 16.75$\pm$0.12 & 3.3$\pm$0.05 & 23.88 & WD (DA) \\
8 & 3412836302516713600\tablefootmark{a} & 5.25$\pm$0.12 & 4.1$\pm$0.05 & 5.74--5.79 & MS (G3) \\
9 & 3853617934132069760 & 6.75$\pm$0.12 & 2.7$\pm$0.05 & 3.52 & MS (M8) \\
10 & 3869392249498967168\tablefootmark{a} & 9.50$\pm$0.12 & 3.2$\pm$0.05 & 4.49--4.89 & MS (F9 - G5) \\
11 & 3992861740235753216\tablefootmark{a} & 11.50$\pm$0.12 & 2.7$\pm$0.05 & 16.09--19.87 & DAZ+M \\
12 & 4437836226304032384 & 13.00$\pm$0.12 & 3.0$\pm$0.05 & 10.04 & DA+M \\
13 & 704748371716051072 & 15.25$\pm$0.12 & 3.3$\pm$0.05 & 3.39 & DAZ+M4 \\ 
14 & 707217153276016000\tablefootmark{a} & 14.75$\pm$0.12 & 3.5$\pm$0.05 & 3.50--3.54 & MS(M2) \\
15 & 830644063706298496 & 24.00$\pm$0.5 & 2.7$\pm$0.05 & 29.90 & DAZ+M  \\
16 & 881086019353249280 & 11.75$\pm$0.12 & 3.1$\pm$0.05 & 14.54 &  DAZ+M \\
\hline

  \end{tabular}
  \tablefoot{The table presents a comparison for WD and MS temperatures estimated in our study with the combined temperature of WD and MS as mentioned in the LAMOST DR11 catalog. Spectral classification from LAMOST catalog are also listed. This is truncated list of cross-matched sources, where we noticed the presence of H-absorption line and/or excess UV emissions. A full cross-matched catalog is available on on \href{https://zenodo.org/records/19270085}{Zenodo}, along with the spectra.  \\
  \tablefoottext{a}{These sources have multi-epoch LAMOST observations.} 
}
\label{tab:lamost_match}
\end{table*}

\begin{table*}
  \caption{A comparison of estimated stellar parameters of WDMS binaries with SIMBAD. }
  \begin{tabular}{|c|c|cc|cc|}
    \hline
  &   &   \multicolumn{2}{c|}{SED parameters}  &  \multicolumn{2}{c|}{SIMBAD parameters }    \\
  \cline{3-6}
 Index &  \gaia\ ID &  WD $\teff$  & MS $\teff$ &  SIMBAD ID   & Spectral Types \\
  &    
  & ($10 ^{3}$K)  & ($10^{3}$K) & 
 ($10 ^{3}$K)  &   \\
    \hline
1 & 1006621281985546240 & 13.50$\pm$0.12 & 3.40$\pm$0.05 & GALEX J063330.3+612325 & D+M3   \\
2 & 1089733400290040320 & 11.00$\pm$0.12 & 3.10$\pm$0.05 & GALEX J073534.3+650649 & DA+M4  \\
3 & 1911816151065187456 & 14.75$\pm$0.12 & 3.30$\pm$0.05 & PM J23283+3319 & DA+dM  \\
4 & 245807579721338752 & 8.00$\pm$0.12 & 2.50$\pm$0.05 & EGGR 570 & DA+dM  \\
5 & 2560009007603950720 & 9.50$\pm$0.12 & 2.70$\pm$0.05 & PHL  3418 & D+dM6  \\
6 & 2864507860881192320 & 12.00$\pm$0.12 & 2.70$\pm$0.05 & GALEX J233910.9+255205 & DA5+M7  \\
7 & 3387904051723656320 & 16.75$\pm$0.12 & 3.30$\pm$0.05 & UCAC4 506-010126 & D+dM4  \\
8 & 3612227169936143360\tablefootmark{a}  & 21.00$\pm$0.5 & 3.40$\pm$0.05 & V* QS Vir & DA3+dM \\
9 & 3662988495054111232 & 8.00$\pm$0.12 & 3.20$\pm$0.05 & LP  618-14 & DA+dM  \\
10 & 3945572123083449984 & 6.25$\pm$0.12 & 2.80$\pm$0.05 & LBQS 1216+1628 & DA+M6 \\
11 & 4028597414327329664 & 5.75$\pm$0.12 & 3.00$\pm$0.05 & G 148-36 & DAZ+M  \\
12 & 4333046892662353152 & 14.50$\pm$0.12 & 3.10$\pm$0.05 & GALEX J165737.2-125633 & DA3+M4  \\
13 & 43789772861265792\tablefootmark{a} & 6.75$\pm$0.12 & 4.70$\pm$0.05 & V* V471 Tau & K2V+DA  \\
14 & 4411741894799595776 & 9.25$\pm$0.12 & 2.90$\pm$0.05 & HS 1606+0153 & DA3+dM  \\
15 & 4437836226304032384 & 13.00$\pm$0.12 & 3.00$\pm$0.05 & PM J16171+0530 & DA+dM  \\
16 & 6182278665776280320 & 12.75$\pm$0.12 & 3.20$\pm$0.05 & EC 13198-2849 & DA5+dM  \\
17 & 6216887306090464000 & 12.50$\pm$0.12 & 3.00$\pm$0.05 & GALEX J143830.0-312719 & DA+?  \\
18 & 759601941671398272 & 5.500$\pm$0.12 & 3.00$\pm$0.05 & EGGR 388 & DC+M4.5Ve  \\
19 & 830644063706298496 & 24.00$\pm$0.5 & 2.70$\pm$0.05 & GD 123 & DA+M3  \\
20 & 881086019353249280 & 11.75$\pm$0.12 & 3.10$\pm$0.05 & GALEX J075919.4+321948 & DA+M  \\
21 & 917083205411949056 & 11.50$\pm$0.12 & 3.30$\pm$0.05 & UCAC4 670-053833 & D+dM3  \\
\hline

  \end{tabular}
  \tablefoot{The table presents a comparison for WD and MS temperatures estimated in our study with the spectral classification of SIMBAD. This is truncated list of cross-matched sources, classified as DA+M spectral type binaries in SIMBAD. The comparison shows a well-match between temperatures estimated using SED analysis and SIMBAD classification. A full cross-matched catalog is available on on \href{https://zenodo.org/records/19270085}{Zenodo}. \\
\tablefoottext{a}{These are classified as spectroscopic binaries. }
}
\label{tab:simbad_match}
\end{table*}

\end{document}